\documentclass[12pt,fleqn,a4paper]{article}
\usepackage[font={small}]{caption}
\usepackage[T1]{fontenc}
\usepackage[utf8]{inputenc}
\usepackage[left=2.5cm,right=2.5cm,top=2.5cm,bottom=2.5cm,foot=0.8cm,]{geometry}
\usepackage{authblk}
\usepackage{hyperref}
\usepackage{amsmath}
\usepackage{amssymb}
\usepackage{amsthm}
\usepackage{bbm}
\usepackage{tabularx}
\usepackage{graphicx}
\usepackage{xcolor}
\usepackage{subcaption}
\usepackage{pdflscape}
\usepackage{booktabs}
\usepackage{natbib}
\usepackage{enumitem}
\newtheorem{proposition}{Proposition}

\theoremstyle{definition} 
\newtheorem{remark}{Remark}

\begin{document}

\begin{titlepage}
\title{Self-respecting worker in the precarious gig economy: A dynamic principal-agent model
}

\author{Zsolt Bihary\thanks{Institute of Finance, Corvinus University of Budapest. E-mail: \href{mailto:zsolt.bihary@uni-corvinus.hu }{zsolt.bihary@uni-corvinus.hu}},~~P\'eter Cs\'oka\thanks{Institute of Finance, Corvinus University of Budapest, and ELTE Centre for Economic and Regional Studies. E-mail: \href{mailto:peter.csoka@uni-corvinus.hu}{peter.csoka@uni-corvinus.hu}},~~P\'eter Ker\'enyi\thanks{Institute of Finance, Corvinus University of Budapest. E-mail: \href{mailto:peter.kerenyi@uni-corvinus.hu }{peter.kerenyi@uni-corvinus.hu}}~~and Alexander Szimayer\thanks{Faculty of Business, Economics and Social Sciences, University of Hamburg.  E-mail: \href{mailto:alexander.szimayer@uni-hamburg.de}{alexander.szimayer@uni-hamburg.de}}}

% \author[1]{Zsolt Bihary}
% \author[1,2]{Péter Csóka}
% \author[1]{Péter Kerényi}
% \author[3]{Alexander Szimayer}
% \affil[1]{Corvinus University of Budapest}
% \affil[2]{Centre for Economic and Regional Studies}
% \affil[3]{University of Hamburg}

%\date{Aug 24, 2022}
% author: inkább preziben, itt más kell
\maketitle
\vspace{-0.3 cm}
\begin{abstract}

We develop a continuous-time principal-agent model of gig work, where contractual flexibility allows the employer to adjust fixed pay and output-based incentives dynamically. The worker’s participation depends on a backward-looking reference value—an exponentially weighted average of net payoff from past employment—capturing a self-respect–driven wage demand. If an offer’s expected utility falls short of the reference value, the worker rejects the contract, resulting in temporary unemployment. Accepted contracts transmit output volatility to the worker through a sensitivity parameter, inducing instability into earnings and employment.  The employer’s optimal threshold policies strategically use this volatility transfer to regulate the worker's wage demands. In the first-best case, the principal imposes maximal volatility to drive the reference value down to its minimum, after which volatility transfer ceases entirely. In the second-best case, the need to incentivize effort ensures ongoing volatility transfer. Stationary analysis predicts a humped-shaped relationship between worker sensitivity and the principal's profit. If sensitivity is too low, the worker's wage demands are overly rigid and difficult to regulate, while if it is too high, their demands become excessively volatile, which eventually erodes the principal's control. 
\vspace{0.15cm} %\\

\noindent
\textbf{Keywords:} Gig economy, Behavioral contract theory, Dynamic contracts, Continuous-time principal-agent model, Business risk transfer, Job rationing, Voluntary unemployment
\vspace{0.15cm}\\
\noindent \textbf{JEL Classification:} C61, D86, D91, J41, J64 \vspace{0.15cm}\\
\end{abstract}

\end{titlepage}

\section{Introduction}

The gig economy—a broad term encompassing contingent, freelance, and alternative work arrangements—refers to labor relationships based on short-term, contract-based projects or ‘gigs’. We use the term ‘gig workers’, covering platform-based food delivery couriers, temporary agency workers in agriculture or manufacturing, self-employed performers, freelance digital nomads, and even fixed-term corporate managers with performance-based compensation (see \cite{spreitzer2017alternative}, \cite{broughton2018experiences}). Despite their differences, these workers share structural features: flexible employment arrangements and compensation schemes in which variable, performance-based pay plays a central role. Such arrangements often impose substantial emotional and financial pressures, contributing to an increasingly precarious existence (\cite{ashford2018surviving}).

One of the defining features of the gig economy is \textit{flexibility}—a term that, in practice, translates into contracts that adapt continuously to changing circumstances. Unlike traditional employment, which offers predictable conditions, gig work links both workload and wages to fluctuating supply and demand. Employers expect workers to adjust rapidly to shifts in consumer needs while they can control the workers through the contracts. As \cite{macdonald2019youth} put it more starkly, “from the point of view of these workers, ‘flexibility’ was a euphemism for exploitation” (p. 733). Such short-term, constantly changing contracts shape not only the conditions of work but also the strategic interaction between employers and workers.

We capture the contractual flexibility within a continuous-time dynamic principal-agent framework. The model describes the evolving relationship between a representative employer (the principal, she) and a representative worker (the agent, he), following the spirit of \cite{holmstrom1987aggregation}, \cite{demarzo2006optimal}, \cite{sannikov2008continuous}, and \cite{cvitanic2012contract}. In this setting, the principal repeatedly offers contracts and adjusts their terms dynamically, combining a fixed payment independent of output with a share of output as performance-based (variable) pay. The output process follows a Brownian motion with non-negative drift, where the drift is proportional to the agent’s effort—capturing both the productivity effect of effort and the inherent output uncertainty from demand fluctuations and other shocks.

Our main contribution to the principal-agent literature is to replace the agent's standard outside option-based participation constraint with an internally anchored heuristic decision rule grounded in his lived experience. Drawing on \cite{rawls1971theory}, who describes self-respect as “perhaps the most important primary good” (p. 440), we model self-respect through an internally anchored benchmark, which we call the \textit{reference value}. This reference value represents the worker’s required expected utility from a prospective contract, defined as an exponentially weighted average of his previously realized net payoffs (wage minus effort costs), and serves as the threshold below which he refuses any contract.

To capture the idea that there is a limit to what the worker will concede, we impose an absolute lower bound on the reference value, representing the worker’s irreducible minimum self-respect. Above the absolute minimum level, if an offer falls short of the reference value, the agent rejects it; output, wage, and thus net payoff drop to zero, and the reference value declines over time, which may lower the agent's wage demand and enable re-entry into employment. Such rejections are interpreted as \textit{voluntary unemployment}—not irrational deviations, but strategic pauses aimed at preserving bargaining power. In this way, self-respect functions as both an emotional buffer and a strategic anchor in an environment of persistent instability.

Although we assume that the agent is risk-neutral in instantaneous decision-making, the dynamic structure of the model creates an indirect channel through which output volatility affects him: fluctuations in output feed into the reference value process, influencing future wage demands and long-term outcomes. This mechanism mirrors what \cite{bieber2021risk} identify as a defining feature of the gig economy: a transfer of business risk from firms to workers through flexible contracts.

Our modeling choice relates to several strands of literature. In psychological game theory (\cite{geanakoplos1989psychological}, \cite{battigalli2009dynamic}), the reference value can be interpreted as the principal’s belief about the agent’s minimum acceptable utility, formed via adaptive expectations. In reference-dependent utility theory (\cite{kHoszegi2006model}), it resembles extreme loss aversion below the benchmark. It also connects to habit formation models (\cite{pollak1970habit}, \cite{abel1990asset}), where past consumption shapes current preferences, though here a wage below the reference triggers refusal to work entirely. In contrast to forward-looking expectations in job search models (\cite{mortensen1986job}, \cite{dellavigna2017reference}) or stochastic outside options (\cite{wang2019optimal}), our participation constraint is \textit{backward-looking}, grounded in the agent’s own wage history.

The principal aims to optimize her expected discounted profit over the long run. From the principal’s perspective, determining the optimal strategy requires solving a stochastic control problem with the agent’s reference value as the state variable. In the first-best case, we assume effort is observable and contractible, allowing the principal to prescribe it directly. This scenario reflects environments where technology, such as advanced AI-based monitoring systems or detailed data analytics, makes effort effectively observable. In the second-best case, effort is unobservable, and incentives must be provided through performance-based pay. In both settings, the optimal policy takes a \textit{threshold form}: when the agent’s reference value is below the threshold, a contract is offered and accepted; when above, the principal may strategically withhold employment to manage future wage demands—a form of \textit{job rationing} (\cite{cave1983job}; \cite{bester1989incentive}).

Under the principal's optimal strategies, the reference value process at the threshold exhibits \textit{sticky reflection} (\cite{harrison1981sticky})\footnote{Also seen in \cite{zhu2012optimal}, \cite{piskorski2016optimal}, and \cite{jacobs2021communities}.}.  We can interpret this reflection as the reference value processes spend positive time on the threshold levels, which can be associated with voluntary unemployment. However, the reflection at the lower bound differs in the two cases. In both the first-best and second-best settings, the principal’s primary tool for managing the agent's wage demand is the deliberate transfer of volatility via an output share. When the agent’s reference value is high, the principal applies the maximum allowable volatility to accelerate its decline, steering it toward the minimum possible level. The key difference emerges once this lower bound is reached. In the first-best case, the principal can switch to a zero-share, fixed-wage contract, making the minimum reference value an absorbing state where the agent is employed at a low, stable wage. In the second-best case, however, the moral hazard constraint requires a persistent positive output share to incentivize effort. Consequently, volatility transfer never ceases, and the lower boundary becomes reflecting, causing the reference value to fluctuate continuously between the minimum floor and the upper threshold.
%However, the reflection at the lower bound differs in the two cases. In the first-best case, the principal deliberately transfers volatility to the agent by offering a share in output. This serves to manage wage demands: when the agent’s reference value is high, increased volatility accelerates its decline, steering the system toward the most desirable state from the principal’s perspective—the minimum reference value—as quickly as possible. Once this state is reached, the principal stops transferring volatility and offers only a fixed wage with no share. The lower boundary becomes absorbing: the agent remains employed at a low, stable wage, and his reference value stays constant. The principal can exploit this favorable state indefinitely. In the second-best case, similar volatility-transfer motives apply, but with an important twist. Here, the output share also serves to incentivize effort, given the moral hazard constraint. As a result, even at the lowest reference value, the principal must offer a positive share, implying continued volatility in the reference value. The lower boundary is thus reflecting rather than absorbing: the reference value fluctuates between the absolute minimum value and the upper threshold value instead of remaining fixed.

Finally, analyzing the \textit{stationary (time-invariant) distribution} of the reference value allows us to characterize long-term averages of output, wages, utility, and profits. The results of our proposed model suggest that increasing the agent’s sensitivity to recent wage fluctuations has non-monotonic effects. Initially, greater sensitivity lowers the agent’s average standard of living and increases the principal’s profit. Beyond a certain point, however, the pattern reverses: excessively sensitive agents become harder to control, and the principal’s ability to manage wage demands weakens. This reveals a fundamental trade-off in flexible labor contracting—between exploiting the agent’s sensitivity and maintaining control over his behavior in the long run.

\section{Model setup}
\label{sec:2}
Our model captures the interaction between an employer (the principal, denoted `she') and a worker (the agent, denoted `he') in the labor market of the gig economy. A standard Brownian motion $B=\left\{ B_t,\mathcal{F}_t;t\geq0\right\}$ on the filtered probability space $(\Omega, {\cal F}, {\cal Q})$ drives the noisy accumulated \textit{output} process $X$, which evolves as
\begin{align*}
%\label{eq:dX}
   dX_t = \chi_t \, (a_t \, dt + \sigma \, dB_t),
\end{align*}
where $\sigma > 0$ denotes the \textit{volatility} of the random component, $a_t \in \mathbb{R}^+_0$ is the agent's \textit{effort level} and $\chi_t \in \{0,1\}$ is the \textit{contract indicator} at time $t\geq0$. If the principal offers an acceptable contract, the agent accepts it and $\chi_t = 1$; otherwise, no agreement is reached and $\chi_t = 0$.

We assume that the principal continuously offers contracts to the agent, each specifying an instantaneous wage composed of a linear function of output. The agent’s accumulated \textit{wage} process $W$ then evolves as
\begin{align*}
%\label{eq:dW}
    dW_t = \chi_t \left(s_t \, dX_t + f_t \, dt\right) = \chi_t \left(\sigma \, s_t \, dB_t + \left(s_t \, a_t + f_t \right) dt\right),
\end{align*}
where $s_t \in [0,\overline{s}]$ denotes the share of output, defining the agent’s variable pay, and $f_t \in \mathbb{R}$ is the fixed pay component. The constant $\overline{s} > 0$ represents the maximum possible share of output that can be allocated to the agent as variable pay. It acts as an upper bound on $s_t$, potentially reflecting regulatory constraints or industry norms. Both $s_t$ and $f_t$ are contract terms offered by the principal at time $t \geq 0$.\footnote{We assume that the strategies of principal and agent $s_t$, $f_t$ and $a_t$ are non-anticipative stochastic processes with respect to the information generated by the output process $X_t$.}

We assume that the principal's accumulated \textit{profit} process $P$ is determined by the residual output remaining after paying the agent’s wage:
\begin{align*}
     dP_t = dX_t - dW_t = \chi_t \left( (1-s_t) \, \sigma \, dB_t + \left((1-s_t) \, a_t - f_t \right) \, dt\right).
\end{align*}
The principal is risk-neutral in our model, meaning she cares only about the expected value of each instantaneous profit.  Accordingly, we define the principal’s \textit{instantaneous expected profit} or simply \textit{profit} $p_t$ as
\begin{align}
\label{eq:pt}
    p_t = E\left[\left. dP_t \, \right| \, {\cal F}_{t-}\right] \, / \, dt = \chi_t \left((1-s_t) \, a_t - f_t \right) .
\end{align}

We postpone the further discussion of the principal's preferences and first turn to those of the agent. In our model, the agent's instantaneous utility reflects both the disutility from exerting effort and the properties of his wage—specifically, its expected value. We model the agent’s \textit{cost of effort} using the standard quadratic specification, $\frac{1}{2} \alpha \, a_t^2$, where $\alpha>0$ is a scaling parameter that ensures the cost is expressed in the same monetary units as the wage.

We do not explicitly model intertemporal savings; rather, we assume the agent is hand-to-mouth: his consumption at each moment equals his wage net of effort cost. Accordingly, the agent's accumulated \textit{net payoff} (in monetary terms) evolves according to
\begin{align*}
%\label{eq:dWtilde}
    dU_t = dW_t - \chi_t \, \frac{\alpha \, a_t^2}{2} \, dt = \chi_t \left(\sigma \, s_t \, dB_t + \left(s_t \, a_t + f_t - \frac{1}{2} \,\alpha \, a_t^2 \right) dt\right) .
\end{align*}

If the agent accepts a contract, he receives a fixed pay component and a share of the stochastic output. By offering share-based compensation, the principal effectively transfers part, or potentially even more than 100 \%, of the output volatility to the agent. To isolate and highlight the dynamic effects of regulating the agent's wage demand, we also assume the agent is risk-neutral. % Nonetheless, incorporating risk-averse preferences is straightforward and does not alter the core mechanics of the model. --- PCS: rather to conclusion. I deleted self-respect here as it is not defined yet. I would rather emphasize dynamic expectation management, also in the title
Under risk neutrality, the agent's (instantaneous) \textit{expected utility} $u_t$ is equivalent to his expected net payoff, which we define as\footnote{We can include risk aversion in our setup by introducing costs $-\frac{1}{2} \, \gamma \,s_t^2\,\sigma^2$ in \eqref{eq:u}, for risk aversion parameter $\gamma \ge 0$, see \eqref{eq:ugamma} in the Appendix. 
%- Yes, appendix, let us make sure that the limit gamma=0 is also ok.
}
\begin{align}
    \label{eq:u}
    u_t = E\left[\left. dU_t \, \right| \, {\cal F}_{t-}\right] \, / \, dt
%    - \frac{1}{2}\, \gamma \left(d\tilde{W}_t\right)^2
    =  \chi_t \left( s_t \, a_t + f_t - \frac{1}{2}\, \alpha\,a_t^2 \right).
%    - \frac{1}{2}\, \gamma\,\sigma^2 \,s_t^2 
\end{align}
%where we have omitted the contract indicator $\chi_t$ for notational simplicity.

In our proposed model, the agent's self-respect is reflected in his decision-making behavior: he accepts a contract at time $t$ if the expected utility of the offer, $u_t$, is greater than or equal to his current \textit{reference value} $R_t$. This decision is not irreversible—after rejecting an offer, the agent remains free to accept future contracts if he chooses to do so. A strictly myopic and classically rational agent would accept any contract that offers non-negative expected utility, since the alternative—no contract—results in no work, no output, and thus zero expected utility. In contrast, in our model, the agent adheres to a personal threshold: he only accepts contracts that yield expected utility at least as high as his reference value, which could also be a very high positive number. The principal observes the agent's reference value, and it is common knowledge that the agent is committed to this rule. In this way, the agent's heuristic decision rule functions as a quasi far-sighted strategy—anchored in self-respect—that effectively protects his long-term standard of living.

The main novelty of our paper is that we define the reference value $R_t$ in a backward-looking and endogenous manner such that $R_t$ is given by
\begin{align}
    \label{eq:expRun}
    R_t = \int _{-\infty}^t \kappa \, e ^{-\kappa \, \left(t-z \right)} \, 
    %%%\left(
    dU_z  
    %%% + (1-\chi_t)\, \underline{r}\, dt \right) 
    \, , \text{ for } t \ge 0 ,
\end{align}
where $\kappa \in \mathbb{R}^+$.
This reference value is an exponentially weighted average of previous net payoff values. The parameter $\kappa$ captures how strongly the agent averages past net payoffs, assigning exponentially declining weights to older experiences. As a higher $\kappa$ places more weight on more recently realized net payoffs, we interpret  $\kappa$ as the agent's \textit{sensitivity parameter} to recent net payoff values. This specification embodies the idea that the worker's self-respect is grounded in his lived experience, specifically his recently realized net payoffs, with an adjustment speed that depends on $\kappa$. The reference value $R_t$ can be interpreted as the agent's required expected utility from a prospective contract. Consequently, the agent only accepts a new contract if he can expect to maintain this expected net payoff level.  
%The reference value is given as an exponentially weighted average of previous net payoff values. We call $\kappa$ the agent's \textit{sensitivity parameter} to recent net payoff values. The Sensitivity parameter $\kappa$ captures how strongly the agent averages past net payoffs, with exponentially declining weights. This specification captures the idea that the worker's self-respect is based on lived experience, namely, the recently realized net payoffs. The agent only accepts a new contract if he can expect to maintain this net payoff level.
Formally, now we can write the contract indicator as
\begin{align}
\label{eq:chi}
    \chi _t &= % {\bf 1}_{ p_t \, \geq \, 0} \cdot {\bf 1}_{ u_t \, \geq \, 0} \cdot 
    \mathbbm{1}_{ u_t \ \geq \, R_t} .
\end{align}

\begin{remark}
    We remark that using psychological games (see \cite{geanakoplos1989psychological} and \cite{battigalli2009dynamic}), $R_t$ can also be interpreted as the principal's belief about the agent's minimal utility expectation of a job, based on adaptive expectations. Then, the psychological instantaneous utility of a job is $\tilde{u}_t=u_t-R_t$, and with a zero utility outside option, the contract indicator can be written as
\begin{align*}
%\label{eq:chi}
    \chi _t &= % {\bf 1}_{ p_t \, \geq \, 0} \cdot 
    \mathbbm{1}_{ \tilde{u}_t \, \geq \, 0} \,,
\end{align*}
which practically corresponds to the rearrangement of the contract indicator in Equation~\eqref{eq:chi}.
\end{remark}

The endogenous formulation of the reference value implies that it evolves as a stochastic process, denoted by $R$, which is implicitly driven by the same Brownian motion $B$ that governs the output process $X$. We assume that the reference value process $R$ is subject to a lower bound: the agent’s reference value $R_t$ cannot fall below a fixed threshold $\underline{r}$. This exogenous parameter $\underline{r} \in \mathbb{R}$ represents the minimum admissible level of self-respect. Throughout the paper, we adopt the natural assumption $\underline{r} = 0$.

Differentiating~Equation~\eqref{eq:expRun} and focusing on $R_t > \underline{r}$, the process $R$  follows the SDE
\begin{align*}
    %\label{eq:dR}
    dR_t & = -\kappa \, R_t \, dt + \kappa \, dU_t  
    & = -\kappa \, R_t \, dt + \chi_t \, \kappa \left(\sigma \, s_t \, dB_t + \left(s_t \, a_t + f_t - \frac{1}{2}\, \alpha \, a_t^2  \right) dt \right) 
    , 
\end{align*}
with initial value $R_0 \ge \underline{r}$. Depending on the values of $a_t$, $s_t$, and $f_t$ at the lower boundary $\underline{r}$, the boundary can be either reflecting or absorbing—an issue we analyze in detail in Section~\ref{sec:Sol_princ_problem}.

Distinguishing the cases whether the offered contract is acceptable or not, we can write the SDE as
\begin{align}
    \label{eq:dRcases}
    dR_t = \begin{cases}
            - \kappa  R_t \, dt \, , &\text{for }\chi_t=0  ,\\
            \kappa \left(s_t \, a_t + f_t - \frac{1}{2} \,\alpha \, a_t^2 - R_t\right)  dt + \kappa \, \sigma \, s_t \, dB_t, &\text{for }\chi_t=1.
            \end{cases} 
\end{align}
It can also be seen here that the sensitivity parameter $\kappa$ captures how strongly the agent averages past realizations of net payoff, effectively determining how far back he looks, with exponentially declining weights.

Finally, let's turn to the principal's preferences. The principal aims to optimize her expected discounted profit over the long run. She employs far-sighted strategies to control the agent's net payoff expectation and, consequently, his wage demand. \footnote{From this point forward, we will use ``wage expectation'' and ``wage demand'' interchangeably with ``net payoff expectation''.}
%Finally, let us turn to the principal's preferences. %However, the worker is myopic in the model, but 
The far-sighted principal's performance, depending on the policies $a_t$, $s_t$, and $f_t$ is measured by the present value of her lifetime profit
\begin{align*}
    E_r \left[\rho \int_0^{\infty}  e^{-\rho t} \, (dX_t-dW_t) \right] =E_r \left[  \rho  \int_0^{\infty} e^{-\rho t} \, \chi_t \, \left((1-s_t)\, a_t - f_t \right) dt \right] ,
\end{align*}
where $\rho$ is her subjective discount rate and $E_r$ denotes the expectation conditional on $R_0 = r$. The factor $\rho$ in front of the integrals normalizes total pay-offs to the same scale as flow pay-offs. 

%WE COULD HAVE THE SAME REMARK with KAPPA.

%The principal's problem is to select an instruction $a_t$ for the agent, a fix-wage $f_t$, a share $s_t$, subject to the agent's demands are satisfied, if this is beneficial to do so. 

\section{Solution of the principal's stochastic control problem}
\label{sec:Sol_princ_problem}

Having established the dynamic framework and behavioral assumptions of both the agent and the principal, we now turn to the core of the analysis: solving the principal’s optimization problem. The objective is to determine the optimal contract—defined by the policies $(a_t, s_t, f_t)$—that maximizes the principal’s expected discounted profit, subject to the agent’s contract indicator and the stochastic evolution of the reference value.

The problem takes the form of a stochastic control problem with participation constraints. These constraints—and the principal’s strategy set—depend on the informational environment, particularly whether the agent’s effort is observable.

In our setup, both the principal and the agent observe the reference value $R_t$, which serves as the state variable of the control problem. The principal’s optimal choices at each point in time are functions of this state. We first analyze the benchmark case in which the principal can observe and directly prescribe effort, leading to the first-best solution. We then turn to the second-best setting, where effort is unobservable and must be induced through incentives embedded in the contract.

\subsection{First-best solution}

The first-best solution in principal-agent theory assumes that the principal observes both the output and the agent’s effort. This observability allows the principal to directly prescribe the desired effort level within the contract. Accordingly, the contractual parameters—effort $a_t$, output share $s_t$, and fixed pay $f_t$—are control variables available to the principal. Owing to the time-homogeneous structure of the problem, these control variables depend only on the current level of the agent’s reference value, which serves as the state variable. Hence, we treat them as functions of the reference value $r$: $a(r)$, $s(r)$, and $f(r)$. The contract indicator $\chi(r)$ similarly becomes a function of $r$.

Given this setup, the principal’s first-best value function is expressed as

\begin{align}
\label{eq:V_FB}
    V^{FB}(r) = \max _{a,s,f} E_r \left[ \rho  \int_0^{\infty}  e^{-\rho \, t} \, \chi(r) \, \Big( \big(1-s(r)\big) \, a(r) - f(r)\Big) \, dt \right] ,
\end{align}
with the dynamics described in Equations~\eqref{eq:chi} and~\eqref{eq:dRcases}. A full characterization of the solution is provided in Proposition~\ref{prop:sol_FB}, with the proof given in Appendix~\ref{sec:App}.

\begin{proposition}
\label{prop:sol_FB}
The value function of the first-best solution $V^{FB}$ in \eqref{eq:V_FB} satisfies
\begin{align*}
    0 = \frac{1}{2\,\alpha} - r - V^{FB}(r) +  \frac{\kappa^2\, \sigma^2 \, \overline{s}^2}{2\,\rho} 
    %\max\left( 0, \frac{\partial^2 V^{FB}}{\partial r^2}(r)   \right) \, ,
    \, \frac{\partial^2 V^{FB}}{\partial r^2}(r)   \, ,
\end{align*}
on $(0, \overline{r}^{FB})$ with free boundary $\overline{r}^{FB}$ and boundary conditions
\begin{align*}
    0 & = V^{FB}(0) - \frac{1}{2\,\alpha} \, ,
    \\
    0 & = - \kappa \, \overline{r}^{FB} \, \frac{\partial V^{FB}}{\partial r}(\overline{r}^{FB}) - \rho\,V^{FB}(\overline{r}^{FB})\, ,
    \\
    0 & = \left( V^{FB}(\overline{r}^{FB}) - \left( \frac{1}{2\,\alpha} - \overline{r}^{FB} \right) \right)  \left( -\kappa\,\overline{r}^{FB}\, \frac{\partial^2 V^{FB}}{\partial r^2}(\overline{r}^{FB}) - (\kappa + \rho)\,\frac{\partial V^{FB}}{\partial r}(\overline{r}^{FB}) \right).
\end{align*}
On $(\overline{r}^{FB}, \infty)$, the value function $V^{FB}$ satisfies
\begin{align*}
    0 & = - \kappa \, r\, \frac{\partial V^{FB}}{\partial r}(r) - \rho\,V^{FB}(r)\, ,
    %%% 0 & = - \kappa\,(r-\underline{r}) \frac{\partial V^{FB}}{\partial r}(r) - \rho\,V^{FB}(r)\, ,
\end{align*}
with solution on $[\overline{r}^{FB},\infty)$ given by 
$V^{FB}(r)  = V^{FB}(\overline{r}^{FB}) \, \left( \frac{r}{\overline{r}^{FB}} \right)^{-\frac{\rho}{\kappa}}$. $V^{FB}$
is continuous on $[0, \infty)$ and continuously differentiable on $(0, \infty)$, and, moreover, twice  continuously differentiable on $(0, \infty)$ if $V^{FB}(\overline{r}^{FB}) > \frac{1}{2\,\alpha} - \overline{r}^{FB}$.
On $[0, \overline{r}^{FB})$ a contract is struck and the optimal controls are
% \begin{align*}
%     a^{FB}(r) & = \frac{1}{\alpha}\, ,
%     \\
%     s^{FB}(r) & = \overline{s}\, \mathbbm{1}_{V^{FB}(r) > \frac{1}{2\,\alpha} - r} \,,
%     \\
%     f^{FB}(r) & = \frac{1}{2\,\alpha} + r - \frac{\overline{s}}{\alpha} \, \mathbbm{1}_{V^{FB}(r) > \frac{1}{2\,\alpha} - r}  \,.
% \end{align*}
\begin{equation*}
    a^{FB}(r)  = \frac{1}{\alpha} ,
    \quad 
    s^{FB}(r)  = \overline{s}\, \mathbbm{1}_{V^{FB}(r) > \frac{1}{2\,\alpha} - r} 
    \,\, \text{  and  }\,\,
    f^{FB}(r) = \frac{1}{2\,\alpha} + r - \frac{\overline{s}}{\alpha} \, \mathbbm{1}_{V^{FB}(r) > \frac{1}{2\,\alpha} - r} .
\end{equation*}
On $[\overline{r}^{FB},\infty)$, no contract is struck and $a^{FB}(r) = s^{FB}(r) = f^{FB}(r) = 0$.
\end{proposition}

The second-order ordinary differential equation (ODE) in Proposition~\ref{prop:sol_FB} involves a free boundary. Consequently, three boundary conditions are required to fully characterize its solution. In general, a closed-form solution does not exist. Nevertheless, we can exploit the structure of the ODE, together with the first boundary condition, which specifies the initial value, to write
\begin{eqnarray*}
    V^{FB}(r) & = & \frac{1}{2\,\alpha} - r + \left(\frac{\partial V^{FB}}{\partial r}(0) + 1\right) \frac{e^{\omega \, r} -  e^{-\omega \, r}}{2\,\omega} \, ,
\end{eqnarray*}
on $[0, \overline{r}^{FB}]$, with $\omega = \frac{\sqrt{2\,\rho}}{\kappa\,\sigma\, \overline{s}}$. This is leaves us with two quantities, $\overline{r}^{FB}$ and $\frac{\partial V^{FB}}{\partial r}(0)$ that can be determined using the remaining two boundary conditions. 
The first of the remaining boundary conditions results in $\frac{\partial V^{FB}}{\partial r}(0)$ as a function of the yet to be determined free boundary $\overline{r}^{FB}$
\begin{eqnarray*}
    \frac{\partial V^{FB}}{\partial r}(0) 
    & = & \frac{2\,(\kappa + \rho)\, \overline{r}^{FB} - \frac{\rho}{\alpha} }{\left(\kappa\,\overline{r}^{FB} + \frac{\rho}{\omega}\right)\, e^{\omega \, \overline{r}^{FB}} + \left(\kappa\,\overline{r}^{FB} - \frac{\rho}{\omega}\right)\, e^{-\omega \, \overline{r}^{FB}}} - 1 \, ,
\end{eqnarray*}
and plugging this in the last of the remaining boundary condition, characterizes the free boundary $\overline{r}^{FB}$ as the solution of\footnote{The equation admits a solution. To see this, observe that the fraction on the right hand side is strictly positive. Consequently, $\overline{r}^{FB} > \frac{1}{2\,\alpha} \, \frac{\rho}{\kappa + \rho}$. For this lower bound, the right hand side is zero. And for $\overline{r}^{FB} \nearrow \infty$, it tends to $\infty$. Continuity then gives the existence of a solution.} 
\begin{eqnarray*}
    1
    & = & \omega \, \left( \frac{\kappa + \rho}{\rho}\, \overline{r}^{FB} - \frac{1}{2\,\alpha} \right) \, \frac{\left(1 + \frac{\omega\,\kappa\,\overline{r}^{FB}}{\kappa + \rho}\right)\,e^{\omega\,\overline{r}^{FB}} + \left(1 - \frac{\omega\,\kappa\,\overline{r}^{FB}}{\kappa + \rho}\right)\,e^{-\omega\,\overline{r}^{FB}}}{\left(1 + \frac{\omega\,\kappa\,\overline{r}^{FB}}{\rho}\right)\,e^{\omega\,\overline{r}^{FB}} - \left(1 - \frac{\omega\,\kappa\,\overline{r}^{FB}}{ \rho}\right)\,e^{-\omega\,\overline{r}^{FB}}} .
\end{eqnarray*}

%\textcolor{red}{PCS: THE MATHEMATICAL CONSIDERATIONS AND INTERPRETATIONS MAY BE HERE.
%$The first-best solution is characterized by a second-order ODE on the interval $(\underline{r},\overline{r}^{FB})$, with a free boundary requiring a total of three boundary conditions to determine the solution. The formulation of the general boundary value problem with three boundary conditions in Proposition~\ref{prop:sol_FB} does not suggest a straightforward solution, either analytically or numerically, as these conditions apply to both boundaries. For instance, at the lower bound $\underline{r}$, the first boundary condition determines the value function as $V^{FB}(\underline{r}) = \frac{1}{2\,\alpha} - \underline{r}$, but its first derivative $\frac{\partial V^{FB}}{\partial r}(\underline{r})$ is not directly determined by the remaining boundary conditions. At the upper bound $\overline{r}^{FB}$, the situation is similar. There, the second and third boundary conditions pin down the solution, provided the free boundary $\overline{r}^{FB}$ is known, but the latter is not directly determined.}

The optimal solution translates into a threshold-based decision from the principal's perspective. This decision hinges on an endogenous threshold reference value, denoted by $\overline{r}^{FB}$. If the agent's reference value is below this threshold, it is optimal for the principal to offer a gig contract that satisfies the agent's wage demand\footnote{More precisely, the agent's demand for utility, accounting for both wage and effort cost.} and is thus acceptable to him. Conversely, if the reference value exceeds $\overline{r}^{FB}$, it is not worthwhile for the principal to employ the agent, whose high wage demand would reduce long-term profitability. In that case, the principal offers a contract that she knows will yield the agent an expected utility below his current reference value, making the offer unacceptable to him.

Strictly speaking, the agent decides whether to accept the contract based on his self-respect (represented by his reference value). However, this decision is indirectly controlled by the principal through the design of the contract. Note that the threshold reference value $\overline{r}^{FB}$ does not imply that the principal could not offer an acceptable contract with positive instantaneous profit at or even above $\overline{r}^{FB}$. Rather, it reflects the far-sighted principal's consideration of the dynamics of the reference value. She deliberately denies employment (formally, by offering an unacceptable contract), thereby inducing the agent to revise his wage demands downward.

\begin{figure}
    \centering
    \caption{Examples of reference value dynamics under optimal controls. All trajectories start from different initial values but are generated using the same simulated path of one single standard Brownian motion. The horizontal dotted lines indicate the threshold reference values where the reference value processes are sticky reflected. We can interpret this reflection as the reference value processes spend positive time on the threshold levels, which can be associated with voluntary unemployment. The $\underline{r}=0$ level is an absorbing barrier in the first-best case and a reflecting barrier in the second-best case. The absorbing barrier in the first-best case is the consequence of the bang-bang control in the share.
    \\Parameter values: output volatility $\sigma=0.2$; agent's effort cost $\alpha=0.2$; sensitivity $\kappa=1.0$; principal's discount rate $\rho=0.2$; and maximum output share $\overline{s}=1.0$.}
    \begin{subfigure}[b]{\textwidth}
    \centering
    \caption{First-best}
    \includegraphics[scale=0.47]{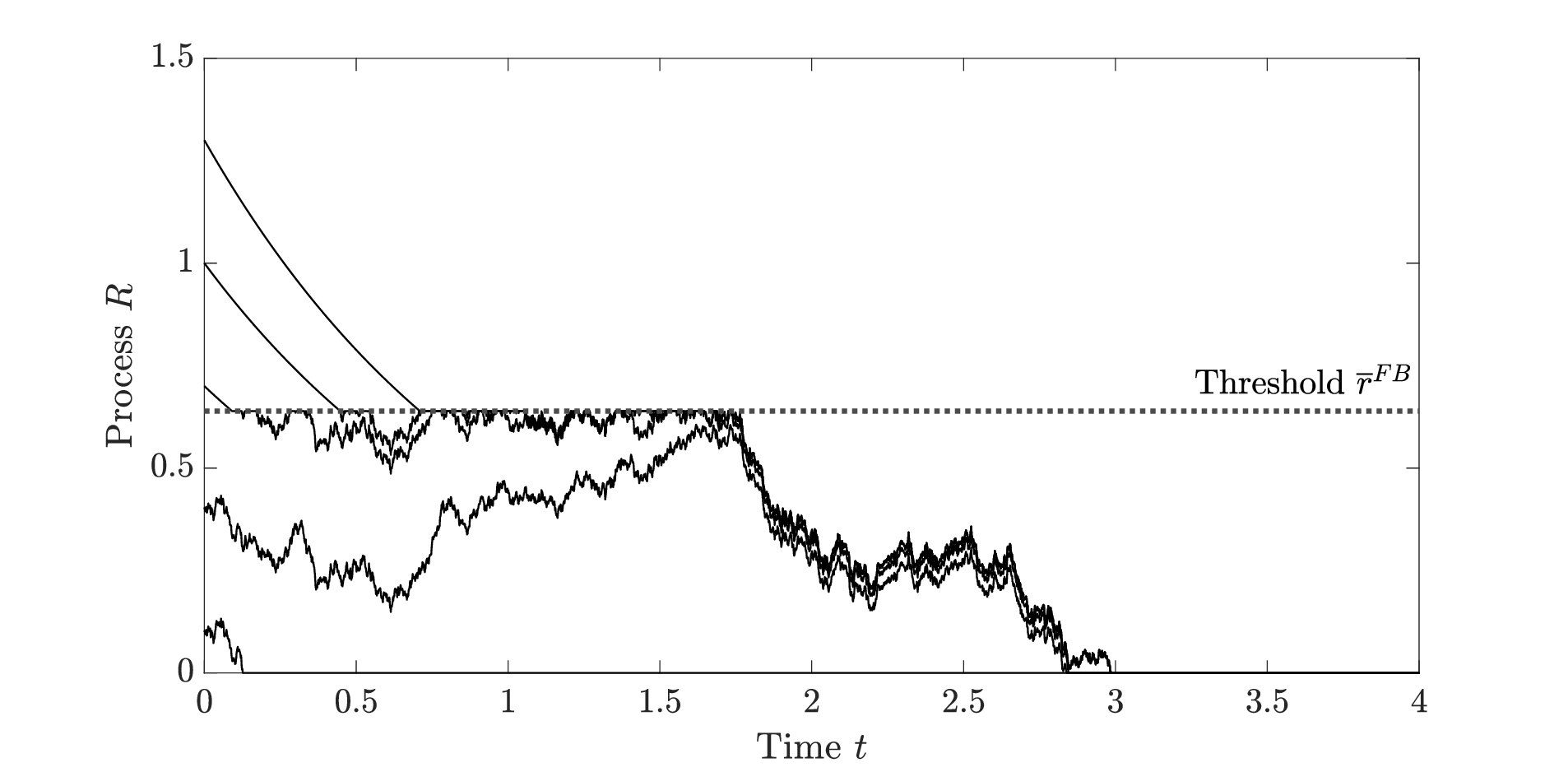}
    \label{fig:Dynamics_FB}
    \end{subfigure}
    \par\bigskip
    \centering
    \begin{subfigure}[b]{\textwidth}
    \centering
    \caption{Second-best}
    \includegraphics[scale=0.47]{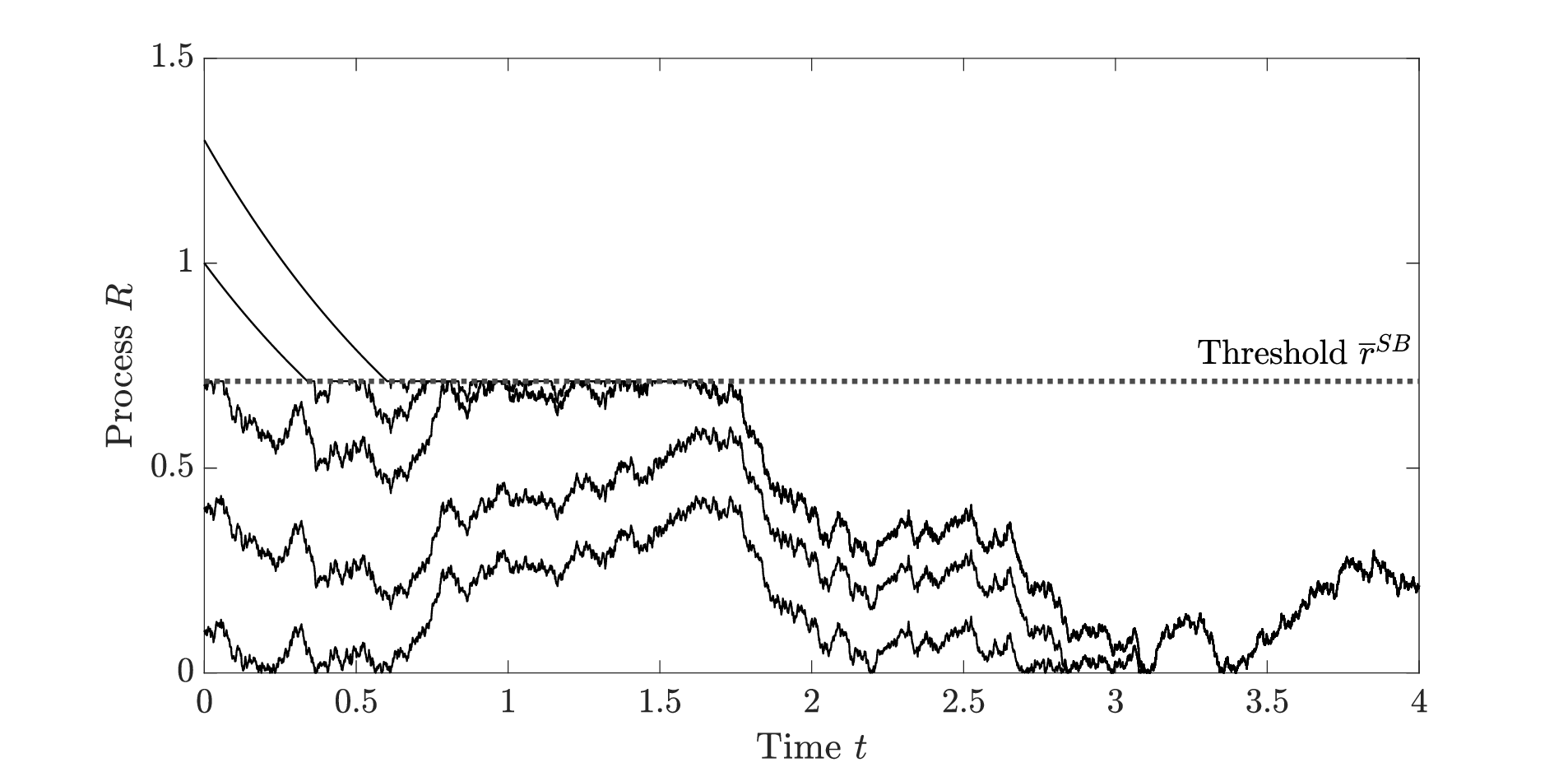}
    \label{fig:Dynamics_SB}
    \end{subfigure}
    \label{fig:Dynamics}
\end{figure}

\begin{figure}
    \centering
    \caption{Principal's value functions and their optimal controls. Dash-dotted and solid lines correspond to the first-best and the second-best solutions, respectively. Vertical dotted lines indicate threshold reference values. On Panel (c), the horizontal dotted line indicates the maximum share $\overline{s}$. Vertical dotted lines and square markers indicate the jumps in optimal share $s^{FB}(r)$ and fixed pay $f^{FB}(r)$ at $r=0$ (bang-bang control) in the first-best case (see Panels (c) and (d)). Optimal effort $a^{SB}(r)$ and share $s^{SB}(r)$ are equal in the second-best case (see solid lines on Panels (b) and (c)).
    \\Parameter values: output volatility $\sigma=0.2$, agent's effort cost $\alpha=0.2$; sensitivity $\kappa=1.0$; principal's discount rate $\rho=0.2$; maximum output share $\overline{s}=1.0$.}
    \label{fig:OptimalSolutions}
    \begin{subfigure}[b]{0.486\textwidth}
    \caption{Principal's value functions}
    \includegraphics[scale=0.47]{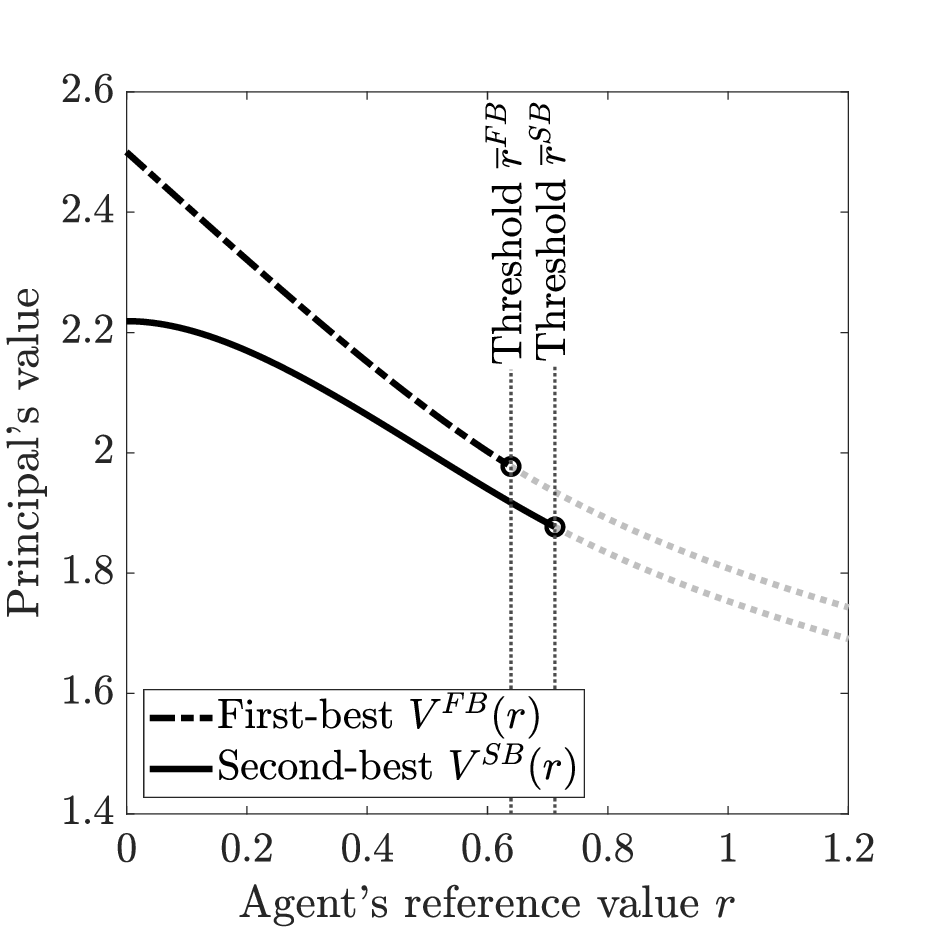}
    \label{fig:OptimalSolutions_valuefunctions}
    \end{subfigure}
    ~
    \begin{subfigure}[b]{0.486\textwidth}
    \caption{Optimal efforts}
    \includegraphics[scale=0.47]{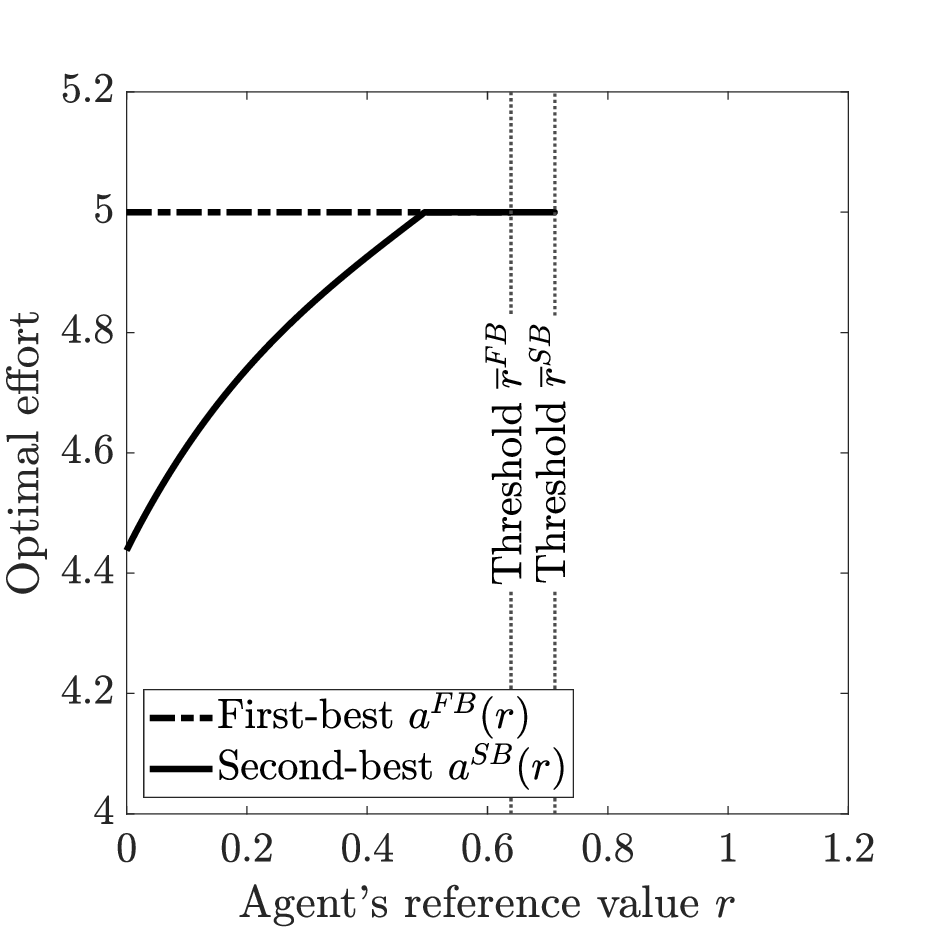}
    \label{fig:OptimalSolutions_efforts}
    \end{subfigure}
    \par\bigskip
    \centering
    \begin{subfigure}[b]{0.486\textwidth}
    \caption{Optimal shares}
    \includegraphics[scale=0.47]{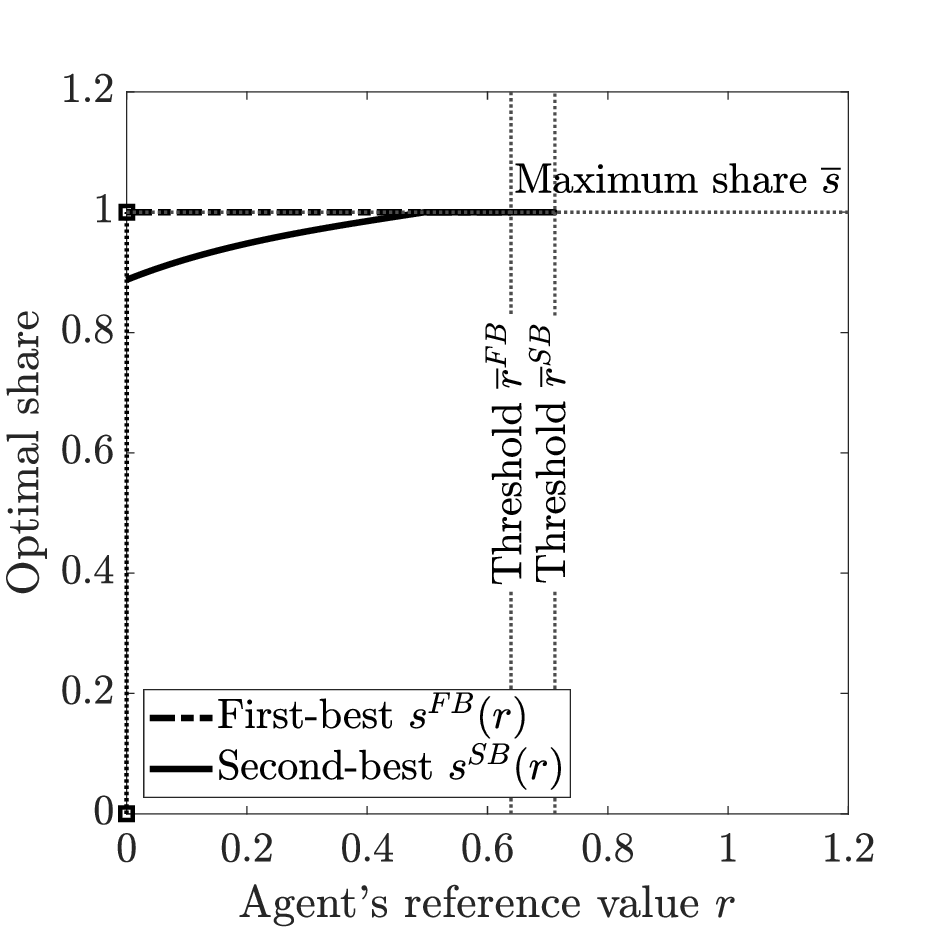}
    \label{fig:OptimalSolutions_shares}
    \end{subfigure}
    ~
    \begin{subfigure}[b]{0.486\textwidth}
    \caption{Optimal fixed pays}
    \includegraphics[scale=0.47]{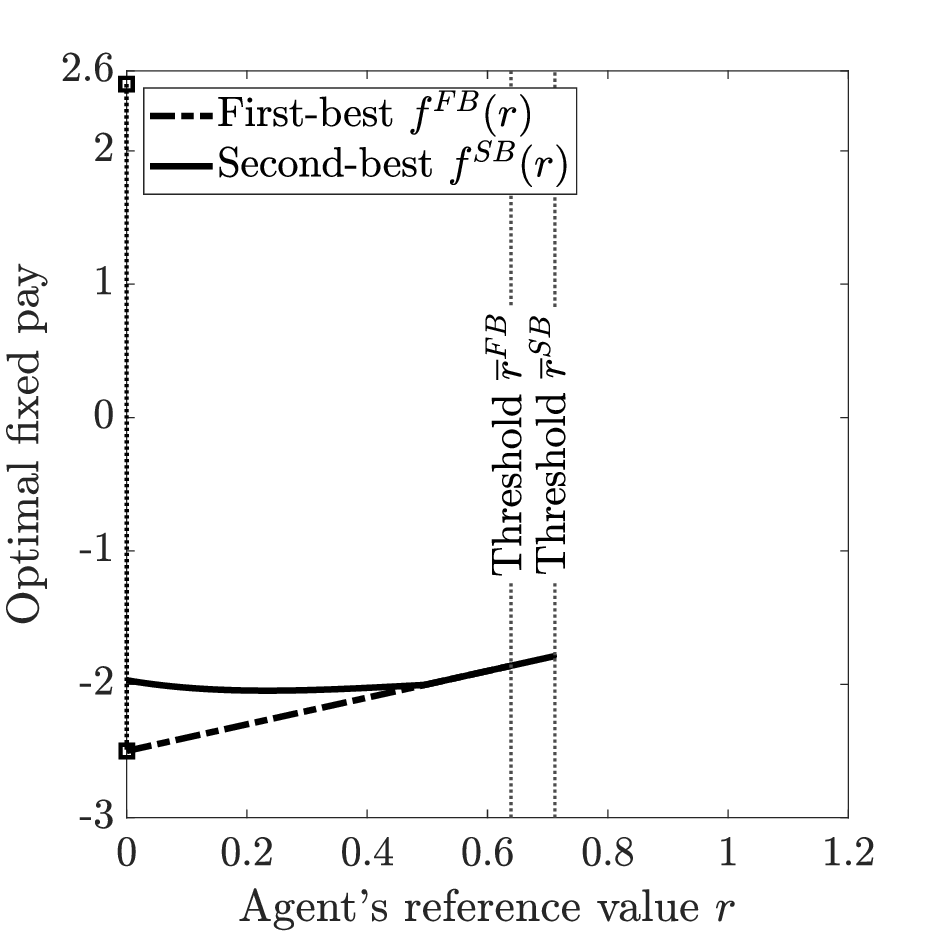}
    \label{fig:OptimalSolutions_fixedpays}
    \end{subfigure}
    
\end{figure}

The contract is only concluded once the agent's reference value falls to the relatively low threshold level $\overline{r}^{FB}$, where the principal can secure higher profits. The threshold reference value, as an endogenously determined upper boundary of the continuous-time reference process, functions as a free boundary with sticky reflection (see Figure \ref{fig:Dynamics_FB}). Of course, the temporarily missing production harms the principal too. However,  the principal performs better in the long run, albeit the cumulative output produced decreases. This behavior represents a specific form of job rationing. From the principal's perspective, the trade-off is between earning relatively low short-term profits by employing an agent with a high wage demand, or accepting a temporary period of zero profits in order to later employ an agent with a lower wage demand and earn higher profits. The principal's value function (see Figure \ref{fig:OptimalSolutions_valuefunctions}) captures the outcome of this trade-off.

In the first-best solution, the principal can directly enforce the agent's effort level, so no incentive mechanism is needed to induce the agent to work optimally. The optimal effort depends solely on the disutility of effort, and the share of output plays no role in determining it. If the principal chooses to employ the agent, the optimal prescribed effort is constant, given by $a^{FB}(r) = 1 / \alpha$ (see Figure \ref{fig:OptimalSolutions_efforts}).

However, the share of output does influence the dynamics of the reference value process $R$. From the principal's perspective, it is optimal if the agent's wage demands are exactly met—that is, if the agent's expected utility $u$ equals his current reference value $r$. Under this condition and the optimal contract parameters, the drift of the reference value process $R$ is zero, and only its volatility depends on the share $s$.

The value $\underline{r} = 0$ is the most desirable state for the principal, as it corresponds to the agent's lowest possible wage demand. Once the reference value reaches this level, the principal sets the share to zero, eliminating both the volatility and the drift of $R$, making $\underline{r} = 0$ an absorbing boundary. The principal can steer the reference value toward this favorable absorbing state by increasing volatility, achieved by offering the maximum share $\overline{s}$ when the agent's reference value is above $\underline{r} = 0$. Thus, the optimal control in the share $s$ takes the form of a bang-bang control: it switches between the minimum value 0 and the maximum value $\overline{s}$, depending on whether the agent's reference value is at the desired lower absorbing boundary $\underline{r} = 0$ or above.

In contrast to the high optimal share, the optimal fixed pay is low—potentially even negative—effectively functioning as a form of rent. Above the absorbing boundary, the principal gives a substantial portion of the output volatility for a fixed rent to the agent. %Even in the first-best case, this contract structure shifts the instantaneous business risk arising from output uncertainty onto the agent. H
% However, even a full output volatility transfer does not fully insulate the principal from the effects of volatility.
% While the agent’s wage is directly affected by output fluctuations, it feeds back into the reference value process, which in turn shapes the agent’s future wage demands. Consequently, output volatility indirectly affects the principal’s long-term profit via the dynamics of the reference value.
However, even a full transfer of output volatility doesn't completely shield the principal from its effects. While the agent's wage is directly impacted by output fluctuations, this in turn feeds into his reference value process, which then shapes his future wage demands. Consequently, output volatility indirectly affects the principal's long-term profit through the dynamics of this reference value.

At the absorbing boundary ($\underline{r} = 0$), the negative fixed pay switches to a positive fixed payment. At this point, the agent receives a relatively low but stable wage and is no longer exposed to output volatility. From this point forward, the principal alone bears all output volatility.

So far, we have examined the principal's first-best solution, in which the principal can prescribe and enforce the agent's effort level. As a result, the effort remains constant and independent of the reference value, and there is no need to incentivize the agent through a share of the output. However, the share still plays a beneficial role for the principal: it increases the volatility of the reference value, enabling the principal to reduce it more quickly to the lowest possible level. Once the reference value reaches this absorbing state, the principal can maintain it indefinitely and continuously exploit the agent's labor at minimal cost.

In what follows, we turn our attention to the second-best solution of the dynamic model. In this setting, the principal can no longer dictate the agent's effort and must instead rely on incentives. This allows us to explore the interplay of incentivization and the dynamic effects of regulating the agent's wage demand.

\subsection{Second-best solution}

In contrast to the first-best case, the principal cannot observe and enforce the agent's effort in the second-best case, therefore, she needs to incentivize him by offering a share of the output. The agent determines his effort $a$ autonomously, optimizing his utility in Equation~\eqref{eq:u}. It follows from this utility maximization that, for a given set of acceptable contract parameters as functions $s$ and $f$, the effort level chosen by the agent is proportional to the share: $a = s / \alpha$. The principal's second-best value function reads as
\begin{align}
\label{eq:V_SB}
    V^{SB}(r) = \max _{s,f} E_r \left[ \rho  \int_0^{\infty}  e^{-\rho\, t} \, \chi(r) \, \left(  \frac{\left( 1-s(r)\right) \, s(r)}{\alpha} - f(r) \,  \right)  {\rm d}t \right],
\end{align}
with the dynamics given in Equations~\eqref{eq:chi} and~\eqref{eq:dRcases}. A characterization of the solution of this control problem is given in Proposition~\ref{prop:sol_SB}, see Appendix~\ref{sec:App} for the corresponding proof.

\begin{proposition}
\label{prop:sol_SB}
Consider the optimal control problem in \eqref{eq:V_SB} with reference value dynamics given by \eqref{eq:dRcases}.
The corresponding value function ${V^{SB}}$ is in $C^2((0,\infty))$ and satisfies the ordinary differential equation %\textcolor{red}{PCS: the following equation seems to have a typo around 0=}
 \begin{eqnarray}
    \nonumber
    0 & = &  \mathbbm{1}_{V^{SB}(r) > \frac{\overline{s}}{2\,\alpha} - r} \left( \frac{\overline{s}}{\alpha}  +\frac{\kappa^2\,\sigma^2}{2\,\rho}\left[  
      \frac{\partial^2 V^{SB}}{\partial r^2}(r) - \frac{\rho}{\alpha\,\kappa^2\,\sigma^2} \right] \overline{s}^2\right)
     \\
\label{eq:ODE_SB}
    & & + \mathbbm{1}_{V^{SB}(r) \le \frac{\overline{s}}{2\,\alpha} -r}  \frac{\frac{\rho}{2\,\alpha^2\,\kappa^2\,\sigma^2}}{\frac{\rho}{\alpha\,\kappa^2\,\sigma^2} - \frac{\partial^2 V^{SB}}{\partial r^2}(r) } 
    - r - V^{SB}(r) \,
 \end{eqnarray}
on $(0, \overline{r}^{SB})$, where $\overline{r}^{SB} > 0$ is a free boundary, with boundary conditions
 \begin{eqnarray*}
    0 & = & \lim_{r \searrow 0}  \frac{\partial V^{SB}}{\partial r}(r)\, ,
    \\
    0 & = & -\kappa\, \overline{r}^{SB} \, \frac{\partial V^{SB}}{\partial r}(\overline{r}^{SB}) - \rho\,V^{SB}(\overline{r}^{SB})\, ,
    %%% 0 & = & \kappa\,(\underline{r} - \overline{r}^{SB}) \frac{\partial V^{SB}}{\partial r}(\overline{r}^{SB}) - \rho\,V^{SB}(\overline{r}^{SB})\, ,
    \\
    0 & = & 
    - \kappa\,\overline{r}^{SB} \, \frac{\partial^2 V^{SB}}{\partial r^2}(\overline{r}^{SB}) - (\rho+\kappa)\,\frac{\partial V^{SB}}{\partial r}(\overline{r}^{SB})\, .
    %%% \kappa\,(\underline{r}-\overline{r}^{SB}) \frac{\partial^2 V^{SB}}{\partial r^2}(\overline{r}^{SB}) - (\rho+\kappa)\,\frac{\partial V^{SB}}{\partial r}(\overline{r}^{SB})\, .
 \end{eqnarray*}
On $(\overline{r}^{SB},\infty)$, the value function satisfies the ordinary differential equation
 \begin{eqnarray*}
% % \nonumber to remove numbering (before each equation)
    0 & = & -\kappa\, r \, \frac{\partial V^{SB}}{\partial r}(r) - \rho\,V^{SB}(r)\, ,
    %%% 0 & = & \kappa\,(\underline{r} - r) \frac{\partial V^{SB}}{\partial r}(r) - \rho\,V^{SB}(r)\, ,
 \end{eqnarray*}
with solution on $[\overline{r}^{SB},\infty)$ given by 
$V^{SB}(r)  = V^{SB}(\overline{r}^{SB}) \, \left( \frac{r }{\overline{r}^{SB}} \right)^{-\frac{\rho}{\kappa}}$.
%%% $V^{SB}(r)  = V^{SB}(\overline{r}^{SB}) \, \left( \frac{r - \underline{r}}{\overline{r}^{SB}-\underline{r}} \right)^{-\frac{\rho}{\kappa}}$. 
On $[0, \overline{r}^{SB})$ a contract is struck and the optimal controls are
\begin{eqnarray*}
   {a}^{SB}(r) & = & \frac{s^{SB}(r)}{\alpha} \,,
   \\
   {s}^{SB}(r)
   & = &  \mathbbm{1}_{V^{SB}(r) \le \frac{\overline{s}}{2\,\alpha} -r} 2 \,\alpha\,(r + V^{SB}(r)) + \mathbbm{1}_{V^{SB}(r) > \frac{\overline{s}}{2\,\alpha} - r}\, \overline{s}
   \,,
    \\
   {f}^{SB}(r) & = & r  - \frac{1}{2\,\alpha} \, s^{SB}(r)^2 \,.
 \end{eqnarray*}
 On $[\overline{r}^{SB},\infty)$, no contract is struck and $a^{SB}(r) = s^{SB}(r) = f^{SB}(r) = 0$.
\end{proposition}

The dynamic-type trade-off between the principal's short-term and long-term perspectives, as captured by the endogenous threshold reference value $\overline{r}$, is also present in this setting. The main difference compared to the first-best solution lies in the role of the share $s$. Since the principal cannot observe or directly prescribe and enforce the agent's effort $a$, she must offer a share to incentivize the agent to exert effort at all. The share $s$ not only affects the volatility of the reference value process but also determines the effort level $a$ and, consequently, the expected instantaneous output $x$ and profit $p$. The principal must offer a share regardless of circumstances, as without it, the expected output would be zero. Therefore, unlike in the first-best case, the principal cannot afford—even at the lower boundary $\underline{r} = 0$—to refrain from offering a share and instead pay only a fixed amount. Even at this boundary, the required share sustains the volatility of the reference value process $R$, making $\underline{r} = 0$ a reflecting barrier (see Figure \ref{fig:Dynamics_SB}).\footnote{The assumption of a lower bound $\underline{r} = 0$ for the reference value is made from the agent’s perspective. Under this assumption, the principal bears no financial exposure to the agent’s behavior at this boundary. Consequently, the boundary condition $\lim_{r \searrow 0}  \frac{\partial V^{SB}}{\partial r}(r) = 0$ in Proposition~\ref{prop:sol_SB} represents a cost-free reflection at $\underline{r} = 0$ from the principal's perspective. However, if the principal becomes involved—such as by preventing the reference value from dropping below $\underline{r} = 0$ (e.g., to avoid agent exit)—the reflection at $\underline{r} = 0$ entails a cost, occurring at a rate of $\frac{\rho}{\kappa}$. In this scenario, the boundary condition is modified to $\lim_{r \searrow 0}  \frac{\partial V^{SB}}{\partial r}(r) = \frac{\rho}{\kappa}$, see Section 3.5 in \cite{dixit1993art} for details.}

The share $s$ increases monotonically with the reference value until it reaches the maximum level $\overline{s}$, which is set as a parameter. A high share $s$ benefits the principal in the short run, as it induces greater effort from the agent, leading to a higher expected instantaneous profit $p$. However, from a dynamic perspective, there is a trade-off associated with a high share. Specifically, higher shares increase the volatility of the reference value process, which can be disadvantageous for the principal when the agent's current wage demand (reference $r$) is low. In such cases, the principal prefers the reference value to remain stable at this low level or even decrease. Since the reference value is bounded below at $\underline{r}=0$, downward movement is limited, whereas upward movement—prompted by high volatility—raises future wage demands, increasing costs for the principal. On the other hand, if the agent's reference value is already relatively high, near the upper threshold $\overline{r}^{SB}$, increased volatility becomes more favorable. In that region, the upward movement of the reference value is constrained by the sticky boundary at $\overline{r}^{SB}$, while the potential for a drop benefits the principal by reducing the agent's future wage demands.

\section{Stationary distributions and average outcomes in the long run}

The endogenous threshold reference value $\overline{r}$ characterizes the maximum instantaneous expected utility $u$ that the agent can attain within the model. In this section, however, our attention shifts from the best-case scenario for to the long-term average outcomes for both the principal and the agent. Specifically, we ask: How do the principal and agent perform on average in the long run? And how does the agent’s sensitivity parameter $\kappa$ influence these aggregate outcomes?

Mathematically, long-term averages are computed by integrating key variables—such as output, profit, and wage—with respect to the stationary (time-invariant) distribution of the reference value process. These averages reflect the typical performance levels that emerge over time and can be interpreted as indicators of economic well-being. In particular, the agent’s long-term average wage serves as a natural proxy for their standard of living in the gig economy.

Alternatively, the stationary distribution can be interpreted as representing a cross-sectional population of heterogeneous agents who differ only in their current reference values. Under this view, the average quantities correspond to the aggregate output produced, wages earned, and profits realized across this continuum of agents—offering a macroeconomic perspective on the micro-level dynamics of contract design and labor allocation.

\subsection{Stationary distributions}

In dynamic stochastic models, the stationary distribution characterizes the long-term behavior of the underlying state variable—in this case, the agent's reference value $R_t$. It describes the probability with which the process occupies different states over time once transitory effects have dissipated. If the process is ergodic and admits a unique stationary distribution, then time averages converge to expectations with respect to this distribution. Accordingly, the stationary distribution provides a powerful tool for analyzing long-term average outcomes, such as output, profit, and wages, as well as for understanding the structural features of the economic relationship between the principal and the agent.

Recall that under the first-best solution, the reference value process $R$ features a sticky reflecting upper boundary at the threshold reference value $\overline{r}^{FB}$ and an absorbing lower boundary at $\underline{r}=0$. Due to the presence of the absorbing lower boundary, the stationary distribution becomes trivial: in the long run, the process will be absorbed at the state $\underline{r}=0$ with probability one. Once this state is reached, the principal can exploit the agent's labor indefinitely.

In the second-best solution, the stationary distribution is more complex. While the upper boundary remains sticky-reflecting, the lower boundary is no longer absorbing but instead reflecting. The stationary distribution $\pi^{SB}$ of the reference value $R_t$  associated with this setup is stated in Proposition~\ref{prop:stationary_SB}, with the corresponding proof provided in Appendix~\ref{sec:App}.

\begin{proposition}
\label{prop:stationary_SB}
Given the second-best setup in Proposition~\ref{prop:sol_SB}, the stationary distribution $\pi^{SB}$ of the reference value $R_t$ on $[0,\overline{r}^{SB}]$ is given by
\begin{align*}
    \pi^{SB}(dr) & = 
    \frac{  s(r)^{-2} }{ \frac{\kappa \, \sigma^2}{2\, \overline{r}^{SB} } +\int_{0+}^{\overline{r}-}  s(u)^{-2} \,{\rm d}u }\,dr\,    
    , \text{ for all } r \in (0,\overline{r}^{SB})\, ,
    \\
    \pi^{SB}(\{ 0\}) & = 0\, , 
    \quad \text{and} \quad
	\pi^{SB}(\{ \overline{r}^{SB}\})  = \frac{ \frac{\kappa \, \sigma^2}{2\, \overline{r}^{SB} }  }{  \frac{\kappa \, \sigma^2}{2\, \overline{r}^{SB} } +\int_{0+}^{\overline{r}-}  s(u)^{-2} \,{\rm d}u }\,.
\end{align*}
\end{proposition}

\begin{figure}
    \caption{Stationary cumulative distribution of the reference value (second-best). The finite jump at the upper threshold represents the finite probability of voluntary unemployment.
    \\Parameter values: output volatility $\sigma=0.2$; agent's effort cost $\alpha=0.2$; sensitivity $\kappa=1.0$; principal's discount rate $\rho=0.2$; and maximum output share $\overline{s}=1.0$.}
    \centering
    % \begin{subfigure}[b]{0.486\textwidth}
    % \caption{Agent's $\kappa=4.0$}
    % \includegraphics[scale=0.484]{Figs/Figure_3_1_stationaryDistribution_kappa4.eps}
    % \label{fig:Dynamics_SB_norm}
    % \end{subfigure}
    % ~
    %\begin{subfigure}[b]{0.486\textwidth}
    %\caption{Agent's $\kappa=0.8$}
    \includegraphics[scale=0.47]{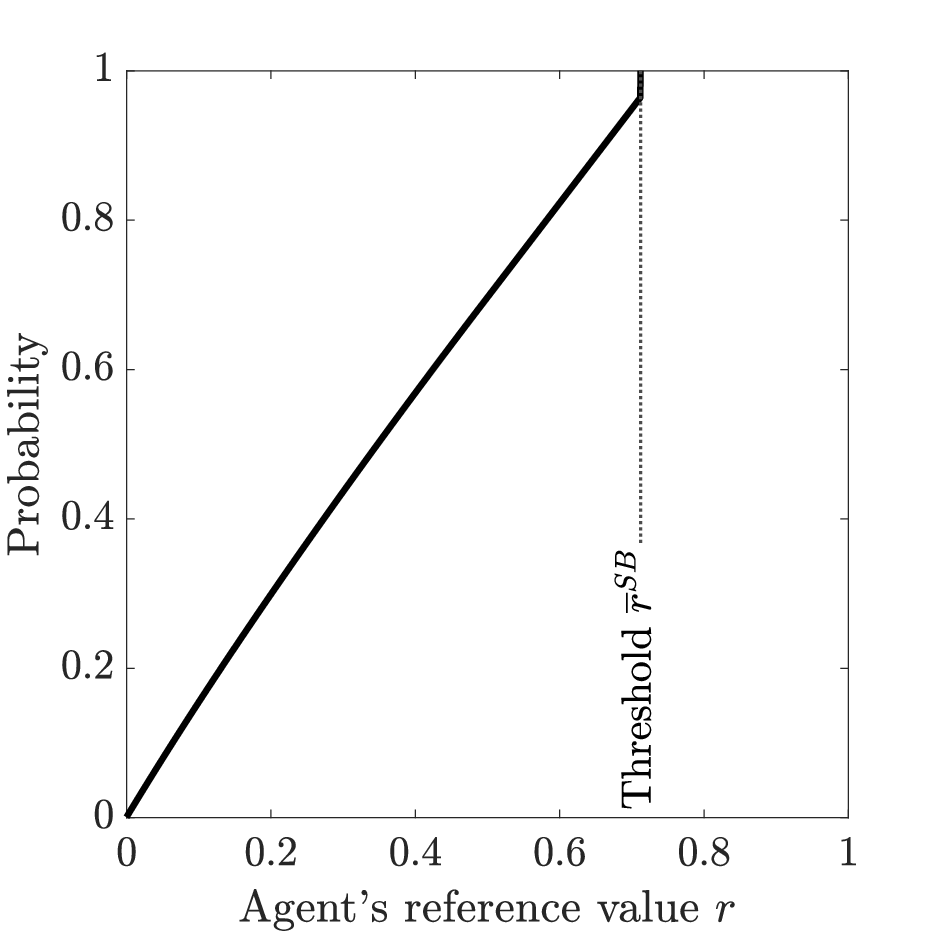}
    \label{fig:stationaryDistribution_SB}
    %\end{subfigure}
    %\label{fig:Dynamics_SB}
\end{figure}

Due to the reflecting nature of the lower boundary $\underline{r} = 0$, the stationary probability at this point is zero. In contrast, the sticky-reflecting behavior at the upper threshold $\overline{r}^{SB}$ leads to a discontinuity in the distribution function. This discontinuity corresponds to the point mass $\pi^{SB}({\overline{r}^{SB}})$, which represents the proportion of time the agent spends in the state of voluntary unemployment. The stationary distributions of the reference value under the second-best solutions are illustrated in Figure~\ref{fig:stationaryDistribution_SB}.

\subsection{Long-term average outcomes}

%{\color{red} ASZ: Should we also do these computations for the first best solution? Here, we have the trivial stationary distribution $R_\infty = \underline{r} = 0$. From there the quantities $\hat{x}^{FB}=\hat{a}^{FB} = {a}^{FB}(0)$,  ${p}^{FB}$, ${\lambda}^{FB}$  $\hat{w}^{FB}$, $\hat{u}^{FB}$ should follow easily, and could serve as a nice benchmark for the second best.--- Yes, we started to add it to the figure captions.}

Having defined the stationary distribution of the reference value, we now turn to the definition of long-term average quantities that characterize the system and the relationship between the principal and the agent. The first such quantity, which reflects the overall performance of the system or the economy, is the \textit{average output}, defined as
\begin{align*}
    \hat{x} = x(0) \, \pi^{SB}(\{ 0\}) + x(\overline{r}^{SB}) \, \pi^{SB}(\{ \overline{r}^{SB}\}) + \int_{0+}^{\overline{r}^{SB}-} x(r) \, \pi^{SB}(dr) ,
\end{align*}
where $x(r)$ denotes the conditional expected value of the instantaneous output, which, by definition, is equal to the agent's instantaneous expected effort $a(r)$. It follows that the average output is equal to the agent's \textit{average effort} $\hat{a}$ (defined analogously), i.e.,
\begin{align*}
    \hat{x}=\hat{a} .
\end{align*}

The principal and the agent share the average output produced between them. The principal receives the \textit{average profit}, defined as
\begin{align*}
     \hat{p} = p(0) \, \pi^{SB}(\{ 0\}) + p(\overline{r}^{SB}) \, \pi^{SB}(\{ \overline{r}^{SB}\}) + \int_{0+}^{\overline{r}^{SB}-} p(r) \, \pi^{SB}(dr) ,
\end{align*}
where $p(r)$ denotes the expected value of the principal's instantaneous expected profit. The proportion of average profit relative to average output indicates how much of the total output accrues to the principal. Since output depends directly on the agent's effort, this ratio can also be interpreted as the extent to which the principal's profit is derived from the agent's labor. Based on this interpretation, we define the \textit{exploitation rate}, denoted by $\lambda$, as
\begin{align*}
    \lambda = \frac{\hat{p}}{\hat{x}}=\frac{\hat{p}}{\hat{a}}.
\end{align*}
An exploitation rate of $\lambda = 0$ implies that the agent receives the full return of their effort, with the principal earning nothing. As $\lambda$ increases, a larger portion of the output generated by the agent's effort accrues to the principal, indicating a higher degree of exploitation.

The remainder of the average output, after subtracting the principal's profit, constitutes the agent's \textit{average wage}, denoted by
\begin{align*}
    \hat{w} = w(0) \, \pi^{SB}(\{ 0\}) + w(\overline{r}^{SB}) \, \pi^{SB}(\{ \overline{r}^{SB}\}) + \int_{0+}^{\overline{r}^{SB}-} w(r) \, \pi^{SB}(dr) ,
\end{align*}
where $w(r)$ is the conditional expected value of the agent's instantaneous wage. Accounting for the disutility of effort, the agent's  average instantaneous expected utility, hereafter referred to as \textit{average utility} is defined as
\begin{align*}
    \hat{u} = u(0) \, \pi^{SB}(\{ 0\}) +u(\overline{r}^{SB}) \, \pi^{SB}(\{ \overline{r}^{SB}\}) + \int_{0+}^{\overline{r}^{SB}-} u(r) \, \pi^{SB}(dr) \,,
\end{align*}
where $u(r)$ is the agent's instantaneous expected utility. The average utility can be interpreted as the agent's long-term standard of living.

In the stationary state, the positive finite probability mass at the upper threshold reference value—resulting from the sticky reflection—can be interpreted as the average fraction of time the agent remains voluntarily unemployed while gradually lowering his wage demands. We refer to this probability as the \textit{voluntary unemployment rate}, denoted by $\nu$:
\begin{align*}
    \nu = \pi^{SB}({ \overline{r}^{SB}}) .
\end{align*}

\section{Agent's sensitivity and long-term performance}

The agent’s sensitivity parameter $\kappa$ plays a central role in shaping long-term economic outcomes in the model. As discussed earlier, $\kappa$ governs the agent’s sensitivity to recently realized net payoffs and determines the pace at which his reference value adjusts. %In this way, $\kappa$ captures how exposed the agent is to wage instability and how reactive—or even erratic—his behavior becomes in response to short-term wage dynamics. These behavioral features influence the volatility of the reference value process and, through it, the structure of the optimal contract.
In this section, we analyze how varying levels of sensitivity, reflected by changes in $\kappa$, affect the stationary distribution of the reference value and, consequently, the long-term averages of output, profit, wage, and utility. The agent’s instability due to high sensitivity makes his behavior harder to anticipate and control, which can ultimately become unfavorable for the principal. Our analysis highlights how the agent’s sensitivity shapes both his long-term standard of living and the principal’s ability to extract surplus over time.

% In the gig-economy context, the agent's time scale parameter $\kappa$, which we associate with sensitivity, is assumed to take relatively high values. This reflects the short memory of the agent, whose wage demands are shaped by realized incomes over only the past few months or even weeks. Accordingly, in the following analysis, we consider $\kappa$ values ranging from 0.2 to 10.

% The principal's time scale parameter is denoted by $\rho$, which corresponds to his subjective discount rate—or, in financial terms, his required rate of return. A lower value of $\rho$ implies greater patience: the principal is more willing to delay production and wait until the worker's wage demand declines before offering a contract. This strategic patience strengthens the principal's bargaining position, allowing him to extract more value over time. In the context of the gig economy, we assume a benchmark value of $\rho$ around $0.2$, reflecting relatively high short-term return expectations—especially when compared to investments involving traditional long-term employment—consistent with the flexible nature of gig-based production. Nevertheless, this still represents a much greater degree of patience than what is implied by the agent's time scale parameter, where high $\kappa$ values correspond to adjustment periods of only a few months or even weeks.

\begin{figure}
    \caption{Average performance as functions of the agent’s sensitivity parameter $\kappa$ in the second-best case. Panel (a) displays long-term average outcomes from the agent’s perspective: average output, wage, and utility. For comparison, in the first-best case with the same parameters, the average output is $5$ (higher than the second-best counterpart), the average wage is $2.5$ (lower than the second-best counterpart), the average utility is zero (lower than the second-best counterpart).  Panel (b) reports the voluntary unemployment rate $\nu$. In the first-best case the voluntary unemployment is zero. Panel (c) shows the principal’s average profit $\hat{p}$. In the first-best case the average profit is $2.5$. Panel (d) depicts the exploitation rate $\lambda$, defined as average profit relative to average output. In the first-best case the exploitation rate is $0.5$.
    \\Parameter values: output volatility $\sigma = 0.2$; agent's effort cost $\alpha = 0.2$; principal’s discount rate $\rho = 0.2$; and maximum output share $\overline{s} = 1.0$.}
    \centering
    \begin{subfigure}[b]{0.486\textwidth}
    \caption{Agent's average performance}
    \includegraphics[scale=0.47]{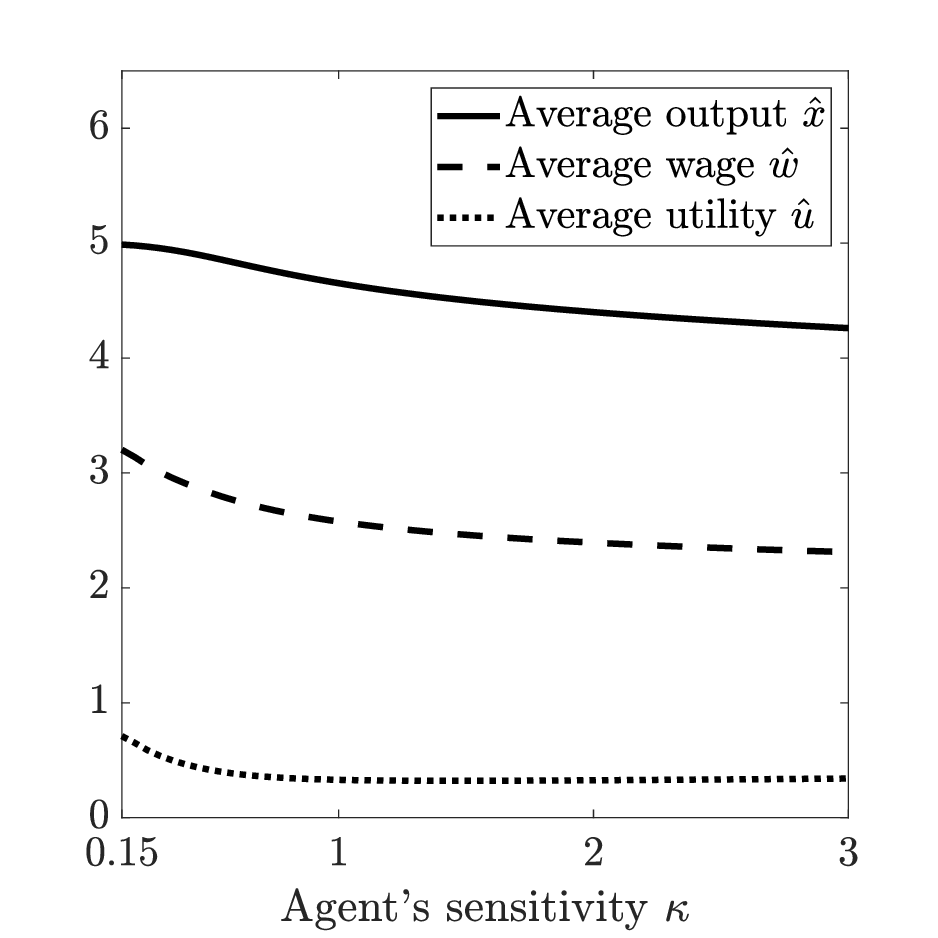}
    \label{fig:Averages_SB_outcomes}
    \end{subfigure}
    ~
    \begin{subfigure}[b]{0.486\textwidth}
    \caption{Voluntary unemployment rate}
    \includegraphics[scale=0.47]{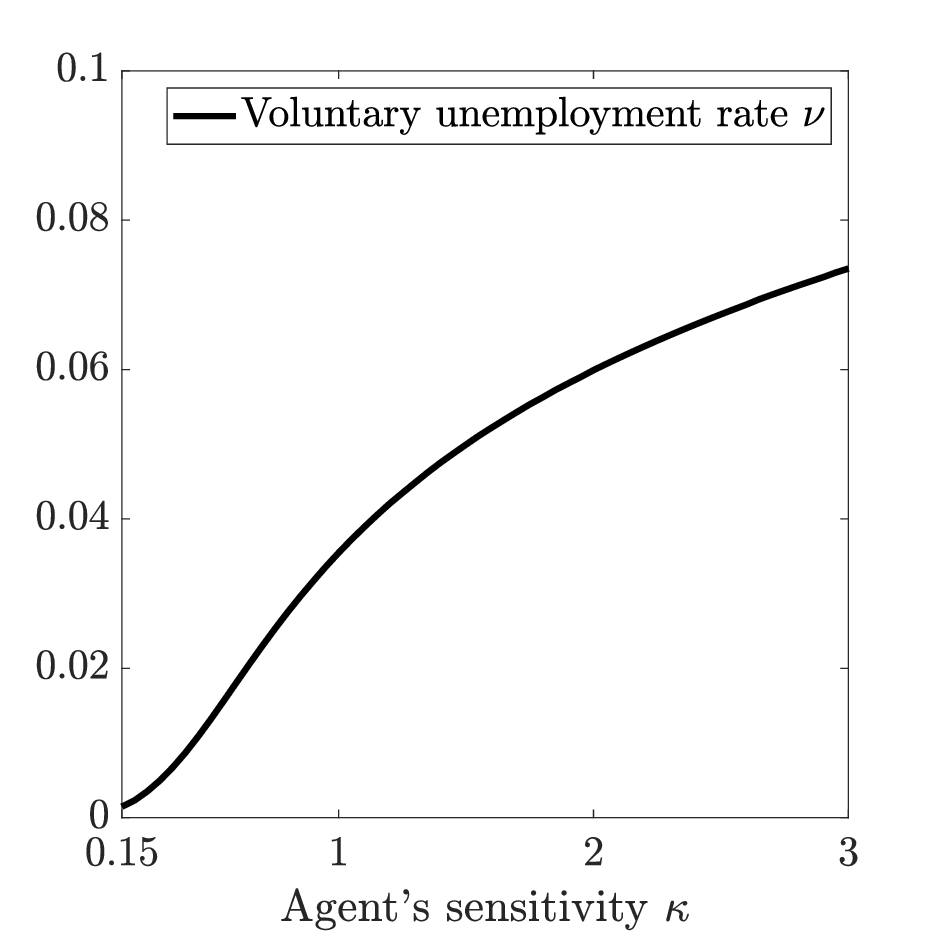}
    \label{fig:Averages_SB_unemployment}
    \end{subfigure}
    \par\bigskip
    \centering
    \begin{subfigure}[b]{0.486\textwidth}
    \caption{Principal's average profit}
    \includegraphics[scale=0.47]{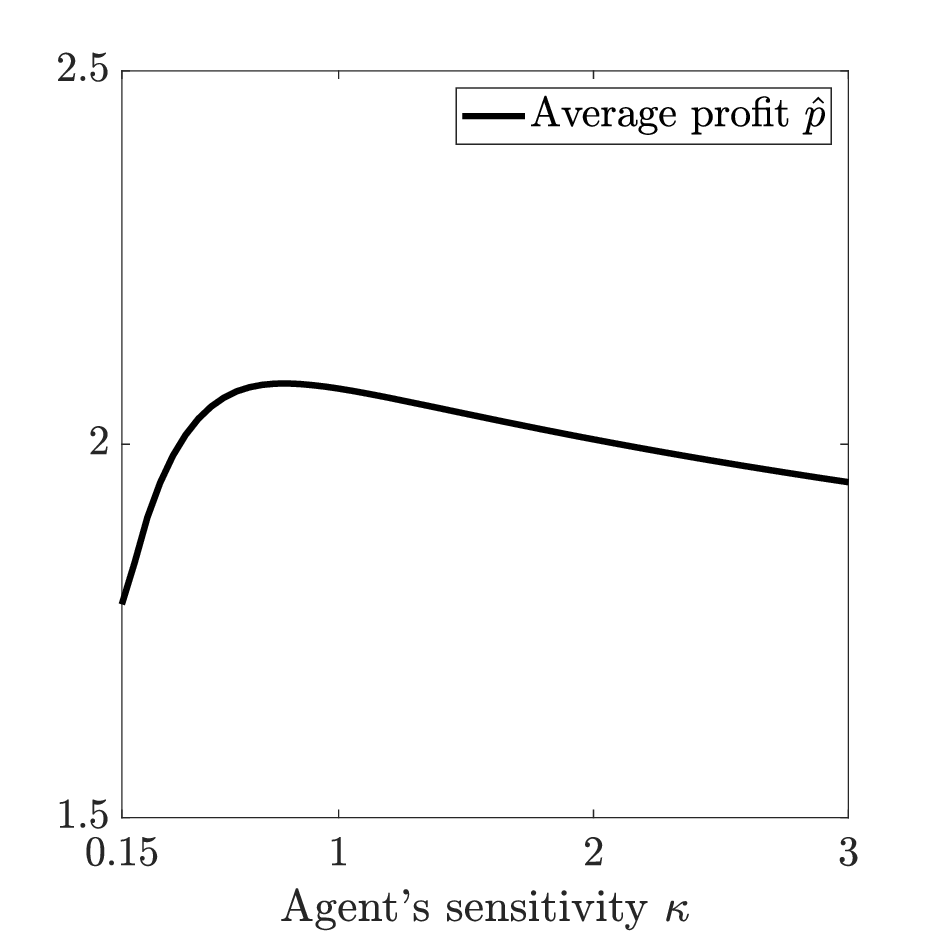}
    \label{fig:Averages_SB_profit}
    \end{subfigure}
    ~
    \begin{subfigure}[b]{0.486\textwidth}
    \caption{Exploitation rate}
    \includegraphics[scale=0.47]{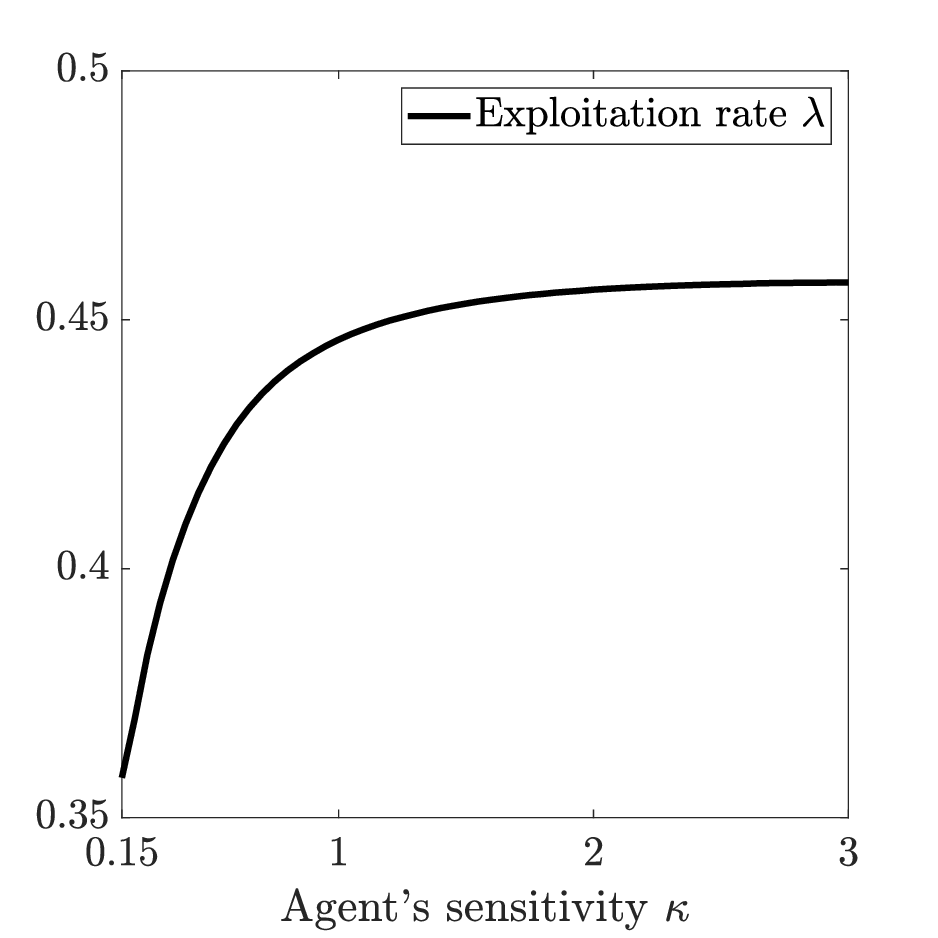}
    \label{fig:Averages_SB_exploitation}
    \end{subfigure}
    \label{fig:Averages_SB}
\end{figure}

Figure~\ref{fig:Averages_SB_outcomes} displays key long-term quantities as functions of the sensitivity parameter $\kappa$ in the second-best case. As sensitivity increases, average output $\hat{x}$ declines monotonically. The more sensitive the agent is to net payoff fluctuations, the less he works on average, leading to lower output. Correspondingly, the average wage $\hat{w}$ also falls.

However, a more nuanced pattern emerges when examining the principal’s average profit $\hat{p}$: after an initial steep increase, the principal’s average profit $\hat{p}$ begins to decline gradually as sensitivity rises further. This suggests that beyond a certain threshold, increased agent sensitivity becomes detrimental to the principal. A similar non-monotonicity appears in the agent’s average (instantaneous expected) utility $\hat{u}$, which first declines and then slowly increases with higher $\kappa$.

The decline in profit is, of course, driven by the reduction in output. However, even when normalized by output—via the exploitation rate, which measures the ratio of profit to output—we observe a similar pattern: after a steep rise, the exploitation rate reaches a peak and then levels off, decreasing only slightly.

To better understand the non-monotonic effect of sensitivity on long-term outcomes, we examine the dynamics of the agent's reference value under optimal second-best control in the contract regime. After substituting the optimal second-best control variables, the relevant form of Equation~\eqref{eq:dRcases} simplifies to:
\begin{align*}
    dR_t = \kappa \sigma s(R_t) \, dB_t.
\end{align*}
Here, for a given level of output volatility $\sigma$, the product of $\kappa$ and the optimal share $s$ determines the volatility of the reference value process. This highlights the amplifying role of $\kappa$: as sensitivity increases, the agent becomes more sensitive to net payoff fluctuations via the stochastic evolution of $R_t$. This increased volatility affects both the frequency of voluntary unemployment and the principal’s ability to stabilize the agent's demands, ultimately shaping the long-term contract structure and outcome distributions.

As discussed in Section~\ref{sec:Sol_princ_problem}, it is desirable for the principal to keep the agent’s reference value low (see the value functions in Figure~\ref{fig:OptimalSolutions}). When the agent’s reference value is high, the principal benefits from increased volatility in the reference value process, as this enhances the likelihood of a downward correction due to the sticky but reflective upper boundary. For a given $\kappa$, this volatility can be increased by raising the agent’s share $s$, as shown in the optimal control.

Conversely, when the reference value is in the lower region, the principal prefers reduced volatility of $R_t$ to prevent upward movement. This can be achieved by offering a lower share. In the first-best case, this strategy is fully effective: at the lowest reference level, the principal sets the share to zero, eliminating volatility entirely. In the second-best case, however, the share $s$ serves a dual purpose: it not only controls volatility but also incentivizes effort. Therefore, it cannot be arbitrarily reduced, as the agent would stop working.

Mathematically, this implies that when $\kappa$ is large, the principal’s ability to reduce the volatility of $R_t$ through $s$ becomes increasingly limited. A highly sensitive agent becomes difficult to control using the share as an instrument. This gives rise to a fundamental trade-off between sensitivity and controllability. Excessively high sensitivity is therefore not in the principal’s interest.
%\textcolor{red}{PCS: sounds like the principal can control sensitivity, but it is just a given property - but maybe she can choose from a pool with different sensitivities?}.
However, this effect remains relatively mild, as reflected by the near-constant exploitation rate even at high levels of $\kappa$.

This trade-off is also evident in the agent’s average utility: after a steep initial decline, it begins to rise gradually. The agent’s extreme sensitivity to recent outcomes—manifested in erratic or `hectic' behavior—reduces the principal’s ability to control him effectively. This unpredictability increases the agent’s implicit bargaining power, despite his sensitivity. 

\section{Conclusion}

In this paper, we developed a continuous-time principal-agent model to explore the complex dynamics of the gig economy, characterized by contractual flexibility and worker precarity. Our central contribution is the introduction of a novel, endogenous, and backward-looking participation constraint. Grounded in the concept of self-respect, this constraint models the worker's wage demand as a reference value determined by an exponentially weighted average of previous net payoff values. In an environment of persistent instability, this psychologically-grounded mechanism acts as both an emotional buffer, shielding the worker’s self-worth, and a strategic anchor for preserving long-term bargaining power.

Our analysis reveals that the principal’s optimal strategy is a threshold policy where she strategically uses periods of unemployment—a form of job rationing—to manage the worker’s wage demands. A key instrument in this process is the transfer of output volatility, capturing business risk. In the first-best case, where effort is observable, the principal applies maximal volatility to drive the worker’s reference value down to its absolute minimum, creating a stable, low-wage absorbing state. In the second-best case, the dual need to incentivize effort and manage wage demands ensures that volatility transfer is perpetual, causing the worker’s reference value to fluctuate continuously between a lower reflecting barrier and the upper threshold. While the main analysis assumes a risk-neutral worker for tractability, we extend the model to include risk aversion in the appendix and show that our core findings remain qualitatively robust.

By examining the stationary distribution of the reference value, we characterized the long-term outcomes for both parties. Our most striking finding is the non-monotonic effect of the worker’s sensitivity ($\kappa$) to recent fluctuations of net payoff values from past employment. While a moderately sensitive worker is easier for the principal to manage and exploit, an excessively sensitive worker becomes unpredictable. This heightened instability erodes the principal's control, leading to lower profits and, counter-intuitively, a slight increase in the agent's long-term average utility. This reveals a fundamental trade-off between the principal’s ability to exploit worker sensitivity and the need to maintain control over the labor relationship.

Ultimately, this paper demonstrates how embedding behavioral realism into dynamic contract theory can yield powerful insights into the precarious nature of modern labor markets. By modeling self-respect as a strategic anchor, we illuminate the subtle interplay of power, flexibility, and control that defines the gig economy.

\section*{Acknowledgments}

We would like to thank Attila Ambrus, Edina Berlinger, János Flesch, Ferenc Horváth, Péter Isztin, Gábor Kertesi, Botond Kőszegi, Sarolta Laczó, Miklós Pintér, Barnabás Szászi, Tibor Takács, Péter Vida, and participants of 9th Annual Financial Market Liquidity Conference, Conference on Mechanism and Institution Design 2020, 4th Workshop on Mechanism Design for Social Good, and Annual Conference of the Hungarian Society of Economics 2023, for helpful comments. Péter Csóka received financial support from the János Bolyai scholarship of the Hungarian Academy of Sciences. Péter Csóka and Péter Kerényi thank funding from National Research, Development and Innovation Office --  NKFIH, K-138826.

\bibliographystyle{plainnat}
\bibliography{references}

\appendix

\section*{Appendix}
\label{sec:App}

Here, we consider a more general formulation of our model by allowing the agent to be risk averse with respect to exposure to the output process. Equation~\eqref{eq:u} is generalized to 
\begin{align}
    \label{eq:ugamma}
    u_t = E\left[\left. dU_t \, \right| \, {\cal F}_{t-}\right] \, / \, dt
    - \frac{1}{2}\, \gamma \, d\tilde{W}_t^2
    =  \chi_t \left( s_t \, a_t + f_t - \frac{1}{2}\, \alpha\,a_t^2 \right)
    - \frac{1}{2}\, \gamma\,\sigma^2 \,s_t^2 ,
\end{align}
where $\gamma \ge 0$ parametrizes the agent's risk aversion. Also, the lower bound $\underline{r}$ is not set to zero, but any non-negative value is allowed, that is, $\underline{r} \ge 0$. 

The extended version of Proposition~\ref{prop:sol_FB} is given by Proposition~\ref{prop:sol_FBgamma} below. The proof of Proposition~\ref{prop:sol_FB} is then a special case of the proof of Proposition~\ref{prop:sol_FBgamma}, setting $\gamma = 0$ and $\underline{r} = 0$.

\begin{proposition}
\label{prop:sol_FBgamma}
For $\underline{r} < \frac{1}{2\,\alpha}$, the value function of the first-best solution $V^{FB}$ is characterized by
\begin{align*}
    0 = \frac{1}{2\,\alpha} - r - V^{FB}(r) +  \frac{\kappa^2\, \sigma^2 \, \overline{s}^2}{2\,\rho} \max\left( 0, \frac{\partial^2 V^{FB}}{\partial r^2}(r)  + \frac{\gamma}{\kappa}\, \frac{\partial V^{FB}}{\partial r}(r)  - \frac{\gamma\,\rho}{\kappa^2}\right) 
\end{align*}
on $(\underline{r}, \overline{r}^{FB})$ with free boundary $\overline{r}^{FB}$ and boundary conditions
\begin{align*}
    0 & = V^{FB}(\underline{r}) - \left( \frac{1}{2\,\alpha} - \underline{r} \right)\, ,
    \\
    0 & = - \kappa \, \overline{r}^{FB} \, \frac{\partial V^{FB}}{\partial r}(\overline{r}^{FB}) - \rho\,V^{FB}(\overline{r}^{FB})\, ,
    \\
    0 & = \left( V^{FB}(\overline{r}^{FB}) - \left( \frac{1}{2\,\alpha} - \overline{r}^{FB} \right) \right)  \left( -\kappa\,\overline{r}^{FB}\, \frac{\partial^2 V^{FB}}{\partial r^2}(\overline{r}^{FB}) - (\rho+\kappa)\,\frac{\partial V^{FB}}{\partial r}(\overline{r}^{FB}) \right).
\end{align*}
On $(\overline{r}^{FB}, \infty)$, the value function $V^{FB}$ satisfies
\begin{align*}
    0 & = - \kappa \, r\, \frac{\partial V^{FB}}{\partial r}(r) - \rho\,V^{FB}(r)\, ,    
\end{align*}
with solution on $[\overline{r}^{FB},\infty)$ given by 
$V^{FB}(r)  = V^{FB}(\overline{r}^{FB}) \, \left( \frac{r}{\overline{r}^{FB}} \right)^{-\frac{\rho}{\kappa}}$. $V^{FB}$
is continuous on $[\underline{r}, \infty)$ and continuously differentiable on $(\underline{r}, \infty)$, and, moreover, twice  continuously differentiable on $(\underline{r}, \infty)$ if $V^{FB}(\overline{r}^{FB}) > \frac{1}{2\,\alpha} - \overline{r}^{FB}$.
On $[\underline{r}, \overline{r}^{FB})$ a contract is struck and the optimal controls are
\begin{align*}
    a^{FB}(r) & = \frac{1}{\alpha}\, ,
    \\
    s^{FB}(r) & = \overline{s}\, \mathbbm{1}_{V^{FB}(r) > \frac{1}{2\,\alpha} - r} \,,
    \\
    f^{FB}(r) & = r - a^{FB}(r)\,s^{FB}(r) + \frac{1}{2} \,\alpha \, a^{FB}(r)^2 + \frac{1}{2}\,\gamma\,\sigma^2 \, s^{FB}(r)^2\,.
\end{align*}
On $[\overline{r}^{FB},\infty)$, no contract is struck and $a^{FB}(r) = s^{FB}(r) = f^{FB}(r) = 0$.
\end{proposition}

\begin{proof}[Proof of Proposition~\ref{prop:sol_FBgamma}]
The principal's objective is to maximize the expected discounted profit net the wage paid to the agent given in Equation~\eqref{eq:V_FB}. In the first-best specification, the principal can observe the agent's effort $a$ and, therefore, $a$ is contractable. Accordingly, the variables $a$, $s$, $f$ are fully controlled by the principal by choosing a suitable admissible contract $(s,f)$. The Hamilton-Jacobi-Bellman (HJB) equation on $(\underline{r},\infty)$ is
\begin{align}
    \nonumber
    0  = & \max_{a,s,f} \kappa\left( \chi \left[a\,s + f - \frac{1}{2}\,\alpha\,a^2  \right]      - r \right) \frac{\partial V^{FB}}{\partial r}(r)  \\
    \label{eq:FB_HJB}
    & \phantom{\max} + \frac{1}{2}\, \chi \,\kappa^2\,\sigma^2 \, s^2 \, \frac{\partial^2 V^{FB}}{\partial r^2}(r) + \rho\,\chi\,\left(a - a\,s - f \right) - \rho\, V^{FB}(r)\,.
\end{align}
Recall that $\chi$ can take only the two values $0$ and $1$ and by Equation~\eqref{eq:chi}
\begin{align}
    \label{eq:FB_pc}
    \chi = 1 
    \quad \Longleftrightarrow \quad a\,s + f -\frac{1}{2}\,\alpha\,a^2 - \frac{1}{2} \,\gamma\,\sigma^2\, s^2 - r \ge 0\,,
\end{align}
describing the agent's participation constraint.
The optimal choice can be determined by maximizing over the two mutually exclusive cases $\chi = 0$ and $\chi = 1$.
For $\chi = 0$, the right hand side of the HJB simplifies to
\begin{align*}
    - \kappa \, r \, \frac{\partial V^{FB}}{\partial r}(r)  - \rho\, V^{FB}(r) \,.
    %%% \kappa\left( \underline{r}- r \right) \frac{\partial V^{FB}}{\partial r}(r)  - \rho\, V^{FB}(r) \,.
\end{align*}
For $\chi = 1$, the right hand side of the HJB reads
\begin{align*}
    \max_{a,s,f \text{ s.t. } \chi = 1} & \kappa\left( \left[a\,s + f - \frac{1}{2}\,\alpha\,a^2  \right] - r \right) \frac{\partial V^{FB}}{\partial r}(r)  
     + \frac{1}{2}\, \kappa^2\,\sigma^2 \, s^2 \, \frac{\partial^2 V^{FB}}{\partial r^2}(r) \\
     & + \rho\,\left(a - a\,s - f \right) - \rho\, V^{FB}(r).
\end{align*}
The objective function depends on $f$ by an additive component that is linear, that is, 
\begin{align*}
    f \,\left( \kappa \,\frac{\partial V^{FB}}{\partial r}(r) - \rho\right)\,.
\end{align*}
To maximize the latter, $f$ needs to be set to its maximal or minimal possible value, depending on the sign of the expression in the bracket.
In the first-best setup, the principal controls $a$, $s$ and $f$ and therefore the value function is non-increasing for increasing wage demand of the agent captured by the reference value, that is, $\frac{\partial V^{FB}}{\partial r} \le 0$.
Thus, the expression in the bracket is strictly negative and the optimal choice for $f$ is to set it to its minimal value given in Equation~\eqref{eq:FB_pc}
\begin{align}
    \label{eq:FB_fopt}
    f^{FB}(a,s) = r - a\,s + \frac{1}{2}\,\alpha\,a^2 + \frac{1}{2} \,\gamma\,\sigma^2\, s^2 \,.
\end{align}
Plugging this in the HJB gives
\begin{align*}
    \max_{a,s} & \, \frac{1}{2} \,\sigma^2  \left(\kappa\,\gamma \,  \frac{\partial V^{FB}}{\partial r}(r)  + \kappa^2\, \frac{\partial^2 V^{FB}}{\partial r^2}(r)  - \rho\,\gamma\right) s^2  + \rho \left(a - \frac{1}{2}\,\alpha\,a^2   - r \right) - \rho\, V^{FB}(r).
\end{align*}
The maximizing values can determined separately
\begin{align*}
    a^{FB} & = \frac{1}{\alpha}\, ,
    \\
    s^{FB} & = \overline{s}\,\mathbbm{1}_{\kappa\,\gamma \,  \frac{\partial V^{FB}}{\partial r}(r)  + \kappa^2\, \frac{\partial^2 V^{FB}}{\partial r^2}(r)  - \rho\,\gamma > 0}\,.
\end{align*}
The right hand side of the HJB for $\chi = 1$ reads then
\begin{align*}
    \frac{1}{2} \,\sigma^2 \, \overline{s}^2 \max\left(0, \kappa\,\gamma \,  \frac{\partial V^{FB}}{\partial r}(r)  + \kappa^2\, \frac{\partial^2 V^{FB}}{\partial r^2}(r)  - \rho\,\gamma\right) + \rho\,\left(\frac{1}{2\,\alpha} - r \right) - \rho\, V^{FB}(r).
\end{align*}
To sum up, the dynamics of the first-best value function $V^{FB}$ has been characterized for the contract regime ($\chi = 1$) and the no contract regime ($\chi = 0$). To proceed, two observations are crucial. First, the principal's maximal profit rate occurs at $R_t = \underline{r}$ and takes the value $\rho\,\left(\frac{1}{2\,\alpha} - \underline{r}\right)$, by setting $a = \frac{1}{\alpha}$, $s = 0$ and $f=\underline{r} + \frac{1}{2\,\alpha}$. The resulting upper bound for the value function is obtained by integrating that value over time using exponential discounting at rate $\rho$, and reads $V^{FB}(r) \le \frac{1}{2\,\alpha} - \underline{r}$, for $r \ge \underline{r}$. At $\underline{r}$ the upper bound is attained in the contract regime with $a^{FB}(\underline{r}) = \frac{1}{\alpha}$, $s^{FB}(\underline{r}) = 0$ and $f^{FB}(\underline{r}) = \underline{r} + \frac{1}{2\,\alpha}$ and 
\begin{align}
    \label{eq:FB_BC1}
    V^{FB}(\underline{r}) & = \frac{1}{2\,\alpha} - \underline{r}\,.
\end{align}
Second, in the no contract regime, the reference point $R$ is strictly decreasing 
$dR_t = - \kappa \, R_t \,dt < 0$, for $R_t >  \underline{r} \ge 0$.
%%% $dR_t = \kappa \,(\underline{r} - R_t) \,dt < 0$, for $R_t > \underline{r}$. 
It follows that a critical threshold $\overline{r}^{FB}$ separates both regimes in such a way that the no contract regime is above the threshold and the contract region is below. Hence, on $(\underline{r}, \overline{r}^{FB})$ it holds
\begin{align}
    \label{eq:FB_ODE}
    0 & = \frac{1}{2\,\alpha} - r - V^{FB}(r) + \frac{\kappa^2\, \sigma^2\, \overline{s}^2}{2\, \rho}   \max\left(0, \frac{\partial^2 V^{FB}}{\partial r^2}(r) + \frac{\gamma}{\kappa} \,  \frac{\partial V^{FB}}{\partial r}(r)     - \frac{\rho\,\gamma}{\kappa^2}\right) .
\end{align}
At the upper boundary $\overline{r}^{FB}$ the value function is continuous. Moreover,  
the value function is continuously differentiable at the upper boundary whenever the reference value process is there not deterministic but stochastic ($s^{FB}(\overline{r}^{FB}) > 0$), resulting in the second boundary condition
\begin{align}
    \label{eq:FB_BC2}
    0 & = 
    - \kappa\,\overline{r}^{FB}\, \frac{\partial V^{FB}}{\partial r}(\overline{r}^{FB}) - \rho\, V^{FB}(\overline{r}^{FB})\,.
    %%% \kappa\,\left(\underline{r} - \overline{r}^{FB}\right)\, \frac{\partial V^{FB}}{\partial r}(\overline{r}^{FB}) - \rho\, V^{FB}(\overline{r}^{FB})\,.
\end{align}
The upper boundary $\bar{r}^{FB}$ is a free boundary and an additional boundary condition to pin down its location is needed.
Applying the optimality of $\bar{r}^{FB}$, the first-order derivative boundary condition can be differentiated to obtain the needed additional boundary condition; it is sometimes called super contact condition, see Sec. 4.6 in \cite{dixit1993art} for a related situation. This gives
\begin{align}
    \label{eq:FB_BC3a}
    0 & = 
    - \kappa\,\overline{r}^{FB}\, \frac{\partial^2 V^{FB}}{\partial r^2}(\overline{r}^{FB}) - (\rho+\kappa)\, \frac{\partial V^{FB}}{\partial r}(\overline{r}^{FB})\,.
    %%% \kappa\,\left(\underline{r} - \overline{r}^{FB}\right)\, \frac{\partial^2 V^{FB}}{\partial r^2}(\overline{r}^{FB}) - (\rho+\kappa)\, \frac{\partial V^{FB}}{\partial r}(\overline{r}^{FB})\,.
\end{align}
According to the lower $V^{FB}(\overline{r}^{FB}) \ge \frac{1}{2\,\alpha} - \overline{r}^{FB}$ (set $a=\frac{1}{\alpha}$, $s =0$ and $f(r) = r + \frac{1}{2\,\alpha}$ on $[\underline{r},\overline{r}^{FB})$) and Equation~\eqref{eq:FB_ODE}, $s^{FB}(\overline{r}^{FB}) > 0$ is equivalent to $V^{FB}(\overline{r}^{FB}) > \frac{1}{2\,\alpha} - \overline{r}^{FB}$, and the third boundary condition is
\begin{align}
    \label{eq:FB_BC3}
    0 & 
    = \left( V^{FB}(\overline{r}^{FB}) -\left( \frac{1}{2\,\alpha} - \overline{r}^{FB}\right) \right) \left( -\kappa\, \overline{r}^{FB}\, \frac{\partial^2 V^{FB}}{\partial r^2}(\overline{r}^{FB}) - (\rho+\kappa) \frac{\partial V^{FB}}{\partial r}(\overline{r}^{FB})\right).
    %%% = \left( V^{FB}(\overline{r}^{FB}) -\left( \frac{1}{2\,\alpha} - \overline{r}^{FB}\right) \right)\left(\kappa\,\left(\underline{r} - \overline{r}^{FB}\right)\, \frac{\partial^2 V^{FB}}{\partial r^2}(\overline{r}^{FB}) - (\rho+\kappa)\, \frac{\partial V^{FB}}{\partial r}(\overline{r}^{FB})\right).
\end{align}
Before, the alternative case $V^{FB}(\overline{r}^{FB}) = \frac{1}{2\,\alpha} - \overline{r}^{FB}$, the value function in the no contract regime $[\overline{r}^{FB}, \infty)$ is explored. There $\chi = 0$ holds and the HJB in Equation~\eqref{eq:FB_HJB} reads
\begin{align*}
    0  & = 
    - \kappa\, r \, \frac{\partial V^{FB}}{\partial r}(r) - \rho\, V^{FB}(r)\,,
    %%% \kappa\, \left(\underline{r}- r \right) \frac{\partial V^{FB}}{\partial r}(r) - \rho\, V^{FB}(r)\,,
\end{align*}
with solution
\begin{align}
    \label{eq:FB_sol0}
    V^{FB}(r) &
    = V^{FB}(\overline{r}^{FB}) \,\left(\frac{r}{\overline{r}^{FB}} \right)^{-\frac{\kappa}{\rho}}\,, \text{ for } r \ge \overline{r}^{FB}\,.
    %%% = V^{FB}(\overline{r}^{FB}) \,\left(\frac{r - \underline{r}}{\overline{r}^{FB}-\underline{r}} \right)^{-\frac{\kappa}{\rho}}\,, \text{ for } r \ge \overline{r}^{FB}\,.
\end{align}
The alternative case $V^{FB}(\overline{r}^{FB}) = \frac{1}{2\,\alpha} - \overline{r}^{FB}$ corresponds to the lower bound $\frac{1}{2\,\alpha} - r$ on $[\underline{r},\overline{r}^{FB}]$ discussed above. In this case $s^{FB} = 0$, the reference value is static ($dR_t = 0$), and the slope of the value function is $-1$, all on $[\underline{r},\overline{r}^{FB})$. 
The optimal upper boundary is determined by first order boundary condition in Equation~\eqref{eq:FB_BC2} matching the slope of the contract and no contract regime at this boundary. This is sufficient, as the value function on the non contract regime is convex. The second order boundary condition in Equation~\eqref{eq:FB_BC3a} does not apply, showing that Equation~\eqref{eq:FB_BC3} covers both cases. Moreover, for the case $s^{FB}=0$ the optimal boundary can be calculated explicitly
\begin{align}
    \label{eq:FB_UpBo}
    \overline{r}^{FB} & 
    =  \frac{\rho}{\rho + \kappa} \, \frac{1}{2\,\alpha}\,.
    %%% = \frac{\kappa}{\rho + \kappa} \, \underline{r} + \frac{\rho}{\rho + \kappa} \, \frac{1}{2\,\alpha}\,.
\end{align}
This finishes the proof.
\end{proof}

Proposition~\ref{prop:sol_SBgamma} below extends Proposition~\ref{prop:sol_SB} in the main part of the paper, as it is allowed for $\gamma \ge 0$ and $\underline{r} \ge 0$. The proof of Proposition~\ref{prop:sol_SB} is then a special case of the proof of Proposition~\ref{prop:sol_SBgamma}, setting $\gamma = 0$ and $\underline{r} = 0$.

\begin{proposition}
\label{prop:sol_SBgamma}
Consider the optimal control problem in \eqref{eq:V_SB} with reference value dynamics given by \eqref{eq:dRcases}.
The corresponding value function ${V^{SB}}$ is in $C^2((\underline{r},\infty))$ and satisfies the ordinary differential equation
 \begin{eqnarray}
    \nonumber
    0 & = &  \mathbbm{1}_{V^{SB}(r) > \frac{\overline{s}}{2\,\alpha} - r} \left( \frac{\overline{s}}{\alpha}  +\frac{\kappa^2\,\sigma^2}{2\,\rho}\left[  
     \frac{\gamma}{\kappa}  \left(\frac{\partial V^{SB}}{\partial r}(r) - \frac{\rho}{\kappa}\right)+ \frac{\partial^2 V^{SB}}{\partial r^2}(r) - \frac{\rho}{\alpha\,\kappa^2\,\sigma^2} \right] \overline{s}^2\right)
     \\
\label{eq:ODE_SBgamma}
    & & + \mathbbm{1}_{V^{SB}(r) \le \frac{\overline{s}}{2\,\alpha} - r}  \frac{\frac{\rho}{2\,\alpha^2\,\kappa^2\,\sigma^2}}{\frac{\rho}{\alpha\,\kappa^2\,\sigma^2} -
    \frac{\gamma}{\kappa}  \left(\frac{\partial V^{SB}}{\partial r}(r) - \frac{\rho}{\kappa}\right) - \frac{\partial^2 V^{SB}}{\partial r^2}(r) } 
    - r - V^{SB}(r) \,.
 \end{eqnarray}
on $(\underline{r}, \overline{r}^{SB})$, where $\overline{r}^{SB} > \underline{r}$ is a free boundary, with boundary conditions
 \begin{eqnarray*}
    0 & = & \lim_{r \searrow \underline{r}}  \frac{\partial V^{SB}}{\partial r}(r)\, ,
    \\
    0 & = & -\kappa\, \overline{r}^{SB} \, \frac{\partial V^{SB}}{\partial r}(\overline{r}^{SB}) - \rho\,V^{SB}(\overline{r}^{SB})\, ,
    %%% 0 & = & \kappa\,(\underline{r} - \overline{r}^{SB}) \frac{\partial V^{SB}}{\partial r}(\overline{r}^{SB}) - \rho\,V^{SB}(\overline{r}^{SB})\, ,
    \\
    0 & = & 
    - \kappa\,\overline{r}^{SB} \, \frac{\partial^2 V^{SB}}{\partial r^2}(\overline{r}^{SB}) - (\rho+\kappa)\,\frac{\partial V^{SB}}{\partial r}(\overline{r}^{SB})\, .
    %%% \kappa\,(\underline{r}-\overline{r}^{SB}) \frac{\partial^2 V^{SB}}{\partial r^2}(\overline{r}^{SB}) - (\rho+\kappa)\,\frac{\partial V^{SB}}{\partial r}(\overline{r}^{SB})\, .
 \end{eqnarray*}
On $(\overline{r}^{SB},\infty)$, the value function satisfies the ordinary differential equation
 \begin{eqnarray*}
% % \nonumber to remove numbering (before each equation)
    0 & = & -\kappa\, r \, \frac{\partial V^{SB}}{\partial r}(r) - \rho\,V^{SB}(r)\, ,
    %%% 0 & = & \kappa\,(\underline{r} - r) \frac{\partial V^{SB}}{\partial r}(r) - \rho\,V^{SB}(r)\, ,
 \end{eqnarray*}
with solution on $[\overline{r}^{SB},\infty)$ given by 
$V^{SB}(r)  = V^{SB}(\overline{r}^{SB}) \, \left( \frac{r }{\overline{r}^{SB}} \right)^{-\frac{\rho}{\kappa}}$.
%%% $V^{SB}(r)  = V^{SB}(\overline{r}^{SB}) \, \left( \frac{r - \underline{r}}{\overline{r}^{SB}-\underline{r}} \right)^{-\frac{\rho}{\kappa}}$. 
On $[\underline{r}, \overline{r}^{SB})$ a contract is struck and the optimal controls are
\begin{eqnarray*}
   {a}^{SB}(r) & = & \frac{s^{SB}(r)}{\alpha} \,,
   \\
   {s}^{SB}(r)
   & = &  \mathbbm{1}_{r+V^{SB}(r) \le \frac{\overline{s}}{2\,\alpha}} 2 \,\alpha\,(r + V^{SB}(r)) + \mathbbm{1}_{r+V^{SB}(r) > \frac{\overline{s}}{2\,\alpha}}\, \overline{s}
   \,,
    \\
   {f}^{SB}(r) & = & r +  
   \left( \frac{1}{2}\,\gamma\,\sigma^2 - \frac{1}{2\,\alpha}\right) s^{SB}(r)^2 \,.
 \end{eqnarray*}
 On $[\overline{r}^{SB},\infty)$, no contract is struck and $a^{SB}(r) = s^{SB}(r) = f^{SB}(r) = 0$.
\end{proposition}

\begin{proof}[Proof of Proposition~\ref{prop:sol_SBgamma}]
The agent maximizes his utility in Equation~\eqref{eq:ugamma} for a given acceptable contract with share of output $s$ and fix wage component $f$. The part of the utility depending on $a$ is $s\,a - \frac{1}{2}\alpha\,a^2$, and the maximizing value is
\begin{align}
    \label{eq:SBa}
    a^{FB}(r) & = \frac{s^{SB}(r)}{\alpha}\,.
\end{align}
The second-best HJB-equation on $(\underline{r},\infty)$ is
\begin{align}
    \nonumber
    0 & = \max_{s,f} \kappa\left( \chi \left[ \frac{s^2}{2\,\alpha} + f \right] 
    %%% + (1-\chi) \underline{r} 
    - r \right) \frac{\partial V^{SB}}{\partial r}(r) + \frac{1}{2}\chi \,\kappa^2\,\sigma^2\, s^2\, \frac{\partial^2 V^{SB}}{\partial r^2}(r)
    \\
    \label{eq:SBHJB}
    & \phantom{ = \max_{s,f} } + \rho\, \chi \left( \frac{s}{\alpha} - \frac{s^2}{\alpha}- f\right) - \rho \, V^{SB}(r)\,. 
\end{align}
As in the first-best case, the objective function depends in the contract regime ($\chi = 1$) on $f$ by an additive component that is linear, that is, 
\begin{align*}
    f \,\left( \kappa \,\frac{\partial V^{SB}}{\partial r}(r) - \rho\right)\,.
\end{align*}
As before, the value function is non-increasing for increasing reference value $r$, that is, $\frac{\partial V^{SB}}{\partial r} \le 0$. Accordingly, the additive component depending on $f$ is maximized for $f$ attaining its minimal value in the contract regime, that is, 
\begin{align}
    \label{eq:SB_fopt}
    f^{SB}(s) = r - \frac{1}{2\,\alpha}\,s^2 + \frac{1}{2} \,\gamma\,\sigma^2\, s^2 \,.
\end{align}
The right hand side of the HJB in Equation~\eqref{eq:SBHJB} in the contract regime reads then
\begin{align*}
    & \max_{0 \le s \le \overline{s}}    \frac{\rho}{\alpha}\, s + \left( - \frac{1}{2}\,\frac{\rho}{\alpha} +
     \frac{1}{2} \,\kappa\, \gamma\,\sigma^2\,  \left[\frac{\partial V^{SB}}{\partial r}(r) - \frac{\rho}{\kappa}\right]+ \frac{1}{2}\kappa^2\,\sigma^2\,  \frac{\partial^2 V^{SB}}{\partial r^2}(r)\right)s^2 
     \\ 
     & \phantom{ \max_{0 \le s \le \overline{s}} } - \rho\, r - \rho \, V^{SB}(r)\,.
\end{align*}
The maximization problem is of the form
\begin{align*}
    \max_{0 \le s \le \overline{s}}  A\,s + B\,s^2 \,,
\end{align*}
with
\begin{align*}
    A = \frac{\rho}{\alpha} \, ,
    \quad \text{ and } \quad B = - \frac{1}{2}\,\frac{\rho}{\alpha} +
     \frac{1}{2} \,\kappa\, \gamma\,\sigma^2\,  \left[\frac{\partial V^{SB}}{\partial r}(r) - \frac{\rho}{\kappa}\right]+ \frac{1}{2}\kappa^2\,\sigma^2\,  \frac{\partial^2 V^{SB}}{\partial r^2}(r)\,.
\end{align*}
As $A > 0$, the optimal share is $s^{SB} = \min(s_0, \overline{s})$ with $s_0 = - \frac{A}{2\, B}$, for $B < 0$, and $s^{SB} = \overline{s}$, for $B \ge 0$. Therefore, the optimal share for the contract regime ($\chi = 1$) is
\begin{align}
    \label{eq:SBs}
    s^{SB}(r) = \begin{cases}
        -\frac{A}{2\, B}\, , & \text{for } B \le -\frac{A}{2\, \overline{s}}
        \\
        \overline{s}\, , & \text{for } B > -\frac{A}{2\, \overline{s}} \,.
    \end{cases}
\end{align}
Suppose $B \le -\frac{A}{2\, \overline{s}}$ and thus $s^{SB} = - \frac{A}{2\, B}$ and $A\, s^{SB} + B\, (s^{SB})^2 = -\frac{A^2}{4\,B}$. The HJB in the contract regime is then
\begin{align*}
    0 & =  -\frac{A^2}{4\,B} - \rho\, r - \rho \,V^{SB}(r)\, .
\end{align*}
Recalling $A = \frac{\rho}{\alpha}$ and solving for $B$ gives
\begin{align*}
    - \frac{\rho}{4\,\alpha^2\,\left( r + V^{SB}(r)\right)} & = B \le - \frac{\rho}{2\, \alpha\, \overline{s}}\, ,
\end{align*}
where the latter inequality holds by assumption. Consequently, the case $B \le - \frac{\rho}{2\, \alpha\, \overline{s}}$ is equivalent to $\frac{\overline{s}}{2\,\alpha} \ge (r+V^{SB}(r))$.
As in the proof of Proposition~\ref{prop:sol_FB}, in the no contract regime, the reference point $R$ is strictly decreasing 
$dR_t = - \kappa \, R_t \,dt < 0$, for $R_t > \underline{r} \ge 0$.
%%% $dR_t = \kappa \,(\underline{r} - R_t) \,dt < 0$, for $R_t > \underline{r}$. 
It follows that a critical threshold $\overline{r}^{SB}$ separates both regimes in such a way that the no contract regime is above the threshold and the contract region is below. Hence, on $(\underline{r}, \overline{r}^{SB})$ it holds
\begin{align}
    \nonumber
    0 & =  \mathbbm{1}_{r+V^{SB}(r) \le \frac{\overline{s}}{2\,\alpha}}  \frac{\frac{\rho}{2\,\alpha^2\,\kappa^2\,\sigma^2}}{\frac{\rho}{\alpha\,\kappa^2\,\sigma^2} -
     \frac{\gamma}{\kappa}  \left[\frac{\partial V^{SB}}{\partial r}(r) - \frac{\rho}{\kappa}\right] - \frac{\partial^2 V^{SB}}{\partial r^2}(r) } 
     \\
     \nonumber
     & + \mathbbm{1}_{r+V^{SB}(r) > \frac{\overline{s}}{2\,\alpha}} \left( \frac{\overline{s}}{\alpha}  +\frac{\kappa^2\,\sigma^2}{2\,\rho}\left[  
     \frac{\gamma}{\kappa}  \left(\frac{\partial V^{SB}}{\partial r}(r) - \frac{\rho}{\kappa}\right)+ \frac{\partial^2 V^{SB}}{\partial r^2}(r) - \frac{\rho}{\alpha\,\kappa^2\,\sigma^2} \right] \overline{s}^2\right)
     \\
     \label{eq:SBODEproof}
      & - r - V^{SB}(r) \,.
\end{align}
The boundary condition at the lower boundary $\underline{r}$ results from $R$ being reflected there and
\begin{align}
    \label{eq:FBBC1}
    0 & = \lim_{r \searrow \underline{r}}  \frac{\partial V^{SB}}{\partial r}(r)\, ,
\end{align}
Following the arguments of the proof of Proposition~\ref{prop:sol_FB}, the boundary conditions at the free boundary $\overline{r}^{SB}$ are
\begin{align}
    \label{eq:FBBC2}
    0 & =
    - \kappa\, \overline{r}^{SB}\, \frac{\partial V^{SB}}{\partial r}(\overline{r}^{SB})
    %%% \kappa\,(\underline{r} - \overline{r}^{SB}) \frac{\partial V^{SB}}{\partial r}(\overline{r}^{SB}) 
    - \rho\,V^{SB}(\overline{r}^{SB})\, ,
    \\
    \label{eq:FBBC3}
    0 & =
    - \kappa\, \overline{r}^{SB} \, \frac{\partial^2 V^{SB}}{\partial r^2}(\overline{r}^{SB})
    %%% \kappa\,(\underline{r}-\overline{r}^{SB}) \frac{\partial^2 V^{SB}}{\partial r^2}(\overline{r}^{SB}) 
    - (\rho+\kappa)\,\frac{\partial V^{SB}}{\partial r}(\overline{r}^{SB})\, .
\end{align}
The value function in the no contract regime $[\overline{r}^{SB}, \infty)$ follows by similar arguments as in the proof of Proposition~\ref{prop:sol_FB} reads
\begin{align*}
    0  & =
    - \kappa\,  r \, \frac{\partial V^{SB}}{\partial r}(r) - \rho\, V^{SB}(r)\,,
    %%% \kappa\, \left(\underline{r}- r \right) \frac{\partial V^{SB}}{\partial r}(r) - \rho\, V^{SB}(r)\,,
\end{align*}
with solution
\begin{align}
    \label{eq:SB_sol0}
    V^{SB}(r) & = V^{FB}(\overline{r}^{SB}) \,
    \left(\frac{r }{\overline{r}^{SB}} \right)^{-\frac{\kappa}{\rho}}
    %%% \left(\frac{r - \underline{r}}{\overline{r}^{SB}-\underline{r}} \right)^{-\frac{\kappa}{\rho}}
    \,, \text{ for } r \ge \overline{r}^{SB}\,.
\end{align}
From the representation above it follows that $V^{SB}$ is twice continuously differentiable on $(\underline{r},\infty)$. For the contract regime, the optimal strategies are
\begin{align*}
    a^{SB}(r) & = \frac{s^{SB}(r)}{\alpha}\, ,
    \\
    s^{SB}(r) & = \mathbbm{1}_{r+V^{SB}(r) \le \frac{\overline{s}}{2\,\alpha}} 2 \,\alpha\,(r + V^{SB}(r)) + \mathbbm{1}_{r+V^{SB}(r) > \frac{\overline{s}}{2\,\alpha}}\, \overline{s}\,,
    \\
    f^{SB}(r) & = r + \frac{1}{2}\left( \gamma\,\sigma^2 - \frac{1}{\alpha} \right)\,s^{SB}(r)^2 \,,
\end{align*}
and the proof is finished.
\end{proof}

Proposition~\ref{prop:stationary_SBgamma} below extends Proposition~\ref{prop:stationary_SB} in the main part of the paper, as it is allowed for $\gamma \ge 0$ and $\underline{r} \ge 0$. The proof of Proposition~\ref{prop:stationary_SB} is then a special case of the proof of Proposition~\ref{prop:stationary_SBgamma}, setting $\gamma = 0$ and $\underline{r} = 0$.

\begin{proposition}
\label{prop:stationary_SBgamma}
Given the second-best setup in Proposition~\ref{prop:sol_SBgamma}, the stationary distribution $\pi^{SB}$ of the reference value $R_t$ on $[\underline{r},\overline{r}^{SB}]$ is given by
\begin{align*}
    \pi^{SB}(dr) & = 
    \frac{ \displaystyle \kappa\, \overline{r}^{SB} \, \frac{2\, e^{- \frac{\gamma}{\kappa}(\overline{r} - r)  } }{\kappa^2\,\sigma^2\, s^2(r)} }{ \displaystyle 1 + \kappa \,\overline{r}^{SB} \, \int_{\underline{r}+}^{\overline{r}-} \frac{2\, e^{- \frac{\gamma}{\kappa}(\overline{r} -u)  }}{\kappa^2\,\sigma^2\, s^2(u)} \,{\rm d}u }\,dr\,
    %%% \frac{ \displaystyle \kappa\,(\overline{r}^{SB}- \underline{r})  \, \frac{2\, e^{- \frac{\gamma}{\kappa}(\overline{r} - r)  } }{\kappa^2\,\sigma^2\, s^2(r)} }{ \displaystyle 1 + \kappa\,(\overline{r}^{SB}- \underline{r}) \, \int_{\underline{r}+}^{\overline{r}-} \frac{2\, e^{- \frac{\gamma}{\kappa}(\overline{r} -u)  }}{\kappa^2\,\sigma^2\, s^2(u)} \,{\rm d}u }\,dr\,
    , \text{ for all } r \in (\underline{r},\overline{r}^{SB})\, ,
    \\
    \pi^{SB}(\{ \underline{r}\}) & = 0\, , 
    \quad \text{and} \quad
	\pi^{SB}(\{ \overline{r}^{SB}\})  = 
    \frac{ 1 }{ \displaystyle 1 + \kappa\,\overline{r}^{SB} \, \int_{\underline{r}+}^{\overline{r}-} \frac{2\, e^{- \frac{\gamma}{\kappa}(\overline{r} - r)  } }{\kappa^2\,\sigma^2\, s^2(r)} \,{\rm d}r } \,.
    %%% \frac{ 1 }{ \displaystyle 1 + \kappa\,(\overline{r}^{SB}- \underline{r}) \, \int_{\underline{r}+}^{\overline{r}-} \frac{2\, e^{- \frac{\gamma}{\kappa}(\overline{r} - r)  } }{\kappa^2\,\sigma^2\, s^2(r)} \,{\rm d}r } \,.
\end{align*}
\end{proposition}

\begin{proof}[Proof of Proposition~\ref{prop:stationary_SBgamma}]
The stationary distribution $\pi^{SB}$ of $R$ on the interval $[\underline {r},\overline{r}^{SB}]$ is determined by the dynamics of $R$ inside the interval and its behavior at the boundaries. The dynamics of $R$ follow
\begin{align*}
    dR_t & = \mu_R(R_t)\,dt + \sigma_R(R_t)\,dB_t
    \\
	& = \frac{1}{2}\,\kappa\,\gamma\,\sigma^2\, s_R(R_t)^2 \, dt + \kappa \, \sigma\, s_R(R_t)\, dB_t \, ,
\end{align*}
At the lower boundary $\underline{r}$, the principal offers with an incentive component, that is, $s^{SB}(\underline{r}) > 0$, and $\sigma_R(\underline{r}) > 0$, thus $R$ is reflected at $\underline{r}$.  
At the upper boundary $\overline{r}^{SB}$, the principal is no longer offering a contract $s^{SB}(\overline{r}^{SB}) = 0$ and $\sigma_R(\overline{r}^{SB}) = 0$. In the no contract regime ($R_t \ge  \overline{r}^{SB}$), the reference value decreases 
$dR_t = - \kappa\, R_t\,dt$,
%%% $dR_t = \kappa(\underline{r} - R_t)dt$, 
thus $R$ is sticky reflected at $\overline{r}^{SB}$.

The stationary distribution $\pi^{SB}$ is characterized by
\begin{align*}
	0 & = \int_{\underline{r}}^{\overline{r}} {\mathcal L} h (r) \,\pi^{SB}(dr) \,,
\end{align*}
where ${\mathcal L}$ is the infinitesimal generator of $R$, for $h \in \mathcal{D}$, see p. 346 in \cite{breiman1992probability}. The infinitesimal generator of $R$ is given by
\begin{align*}
	{\mathcal L}h(r) & = 
	\begin{cases}
		\mu_R(r) \, h^\prime(r) + \frac{1}{2}\,\sigma_R^2(r) \, h^{\prime\prime}(r)\, , & \text{ for } r \in (\underline{r},\overline{r}^{SB}) 
		\\
		\mu_R(\underline{r}+) \, h^\prime(\underline{r}+) + \frac{1}{2}\,\sigma_R^2(\underline{r}+) \, h^{\prime\prime}(\underline{r}+)\, , & \text{ for } r = \underline{r} 
		\\
		\mu_R(\overline{r}^{SB}-) \, h^\prime(\overline{r}^{SB}-) + \frac{1}{2}\,\sigma_R^2(\overline{r}^{SB}-) \, h^{\prime\prime}(\overline{r}^{SB}-)
		\, , & \text{ for } r = \overline{r}^{SB}\,.		
	\end{cases}
\end{align*}
with domain ${\mathcal D} = \{h : h , {\mathcal L} \in  {\mathcal C}_b([\underline{r},\overline{r}^{SB}]), h^\prime(\underline{r}+) = 0,  c\,h^{\prime\prime}(\overline{r}^{SB}-) = h^{\prime}(\overline{r}^{SB}-) \}$ where ${\mathcal C}_b([\underline{r},\overline{r}^{SB}])$ denotes the set of continuous and bounded
functions from $[\underline{r},\overline{r}^{SB}]$ on $\mathbb{R}$, see II.1.7-9 in \cite{borodin2012handbook} for the specification of $\mathcal{L}$ and in particular for the effect of the boundary behavior of $R$ on $\mathcal D$.
Here, the constant $c$ satisfies
\begin{align*}
	\mu_R(\overline{r}^{SB}-)  + \frac{1}{2\, c}\,\sigma_R^2(\overline{r}^{SB}-) & =  
    - \kappa \,  \overline{r}^{SB} \,,
    %%% \kappa \, (\underline{r} - \overline{r}^{SB} )\,,
\end{align*}
hence
\begin{align*}
	\mu_R(\overline{r}^{SB}-) \, h^\prime(\overline{r}^{SB}-) + \frac{1}{2}\,\sigma_R^2(\overline{r}^{SB}-) \, h^{\prime\prime}(\overline{r}^{SB}-)
	= 
    - \kappa\,\overline{r}^{SB} \,h^\prime(\overline{r}^{SB}-) \,.
    %%% \kappa\,(\underline{r}-\overline{r}^{SB}) \,h^\prime(\overline{r}^{SB}-) \,.
\end{align*}
Since $\pi^{SB}(\{\underline{r}\}) = 0$, the stationary distribution satisfies 
\begin{align*}
	0  & =  \int_{\underline{r}}^{\overline{r}^{SB}} {\mathcal L} h (r) \,\pi^{SB}(dr)  
    \\
    & =  \int_{\underline{r}+}^{\overline{r}^{SB}-} \left( \mu_R(r) \, h^\prime(r) + \frac{1}{2}\, \sigma_R^2(r) \, h^{\prime\prime}(r) \right) \pi^{SB}(dr) 
    - \kappa\,\overline{r}^{SB} \, h^\prime(\overline{r}^{SB}-) \, 
    %%% + \kappa\,(\underline{r}-\overline{r}^{SB})  \, h^\prime(\overline{r}^{SB}-) \, 
    \pi^{SB}(\{ \overline{r}^{SB}\}) \, ,
\end{align*}
for all $h \in {\mathcal D}$.
Now, $h^\prime(\overline{r}^{SB}-) = h^\prime(r) + \int_r^{\overline{r}^{SB}-} h^{\prime\prime}(u) \,du$ and 
\begin{align*}
	0  = & \int_{\underline{r}+}^{\overline{r}^{SB}-}  \frac{1}{2}\, \sigma_R^2(r) \, h^{\prime\prime}(r) \,\pi^{SB}(dr) -\int_{\underline{r}+}^{\overline{r}^{SB}-} \int_r^{\overline{r}^{SB}-} \mu_R(r) \, h^{\prime\prime}(u) \,{d}u \,\pi^{SB}(dr)  
	\\ 
	& + h^\prime(\overline{r}^{SB}-) \, \left(\int_{\underline{r}+}^{\overline{r}^{SB}-} \mu_R(r) \, \pi^{SB}({d}r) 
    - \kappa\,\overline{r}^{SB}  \, \pi^{SB}(\{ \overline{r}\})
    %%% + \kappa\,(\underline{r}-\overline{r}^{SB})  \, \pi^{SB}(\{ \overline{r}\}) 
    \right)
	\\
    = & \int_{\underline{r}+}^{\overline{r}^{SB}-}  \frac{1}{2}\, \sigma_R^2(r) \, h^{\prime\prime}(r) \,\pi^{SB}(dr) -\int_{\underline{r}+}^{\overline{r}^{SB}-} \int_{\underline{r}+}^r \mu_R(u) \, h^{\prime\prime}(r) \,\pi^{SB}({d}u) \,dr  
	\\ 
	& + h^\prime(\overline{r}^{SB}-) \, \left(\int_{\underline{r}+}^{\overline{r}^{SB}-} \mu_R(r) \, \pi^{SB}({d}r) 
    - \kappa\,\overline{r}^{SB} \, \pi^{SB}(\{ \overline{r}\}) 
    %%% + \kappa\,(\underline{r}-\overline{r}^{SB})  \, \pi^{SB}(\{ \overline{r}\}) 
    \right)
    \\
	  = & \int_{\underline{r}+}^{\overline{r}^{SB}-} h^{\prime\prime}(r) \, \left( \frac{1}{2}\, \sigma_R^2(r) \, \pi^{SB}({d}r) 
	- \int_{\underline{r}-}^r \mu_R(u)  \,\pi^{SB}({d}u)\,{d}r\right)  
	\\ 
	& + h^\prime(\overline{r}^{SB}-) \, \left(\int_{\underline{r}+}^{\overline{r}^{SB}-} \mu_R(r) \, \pi^{SB}({d}r) 
    - \kappa\,\overline{r}^{SB}
    %%% + \kappa\,(\underline{r}-\overline{r}^{SB})  
    \, \pi^{SB}(\{ \overline{r}^{SB}\}) \right)
\end{align*}
One may assume that the stationary distribution is absolutely continuous with respect to the Lebesgue measure on $(\underline{r},\overline{r}^{SB})$, that is, $\pi^{SB}(dr) = \phi(r)\, dr$, for $\underline{r} < r < \overline{r}^{SB}$. Then, integration by parts gives
\begin{align*}
    0  = & \left. h^{\prime}(r) \, \left( \frac{1}{2}\, \sigma_R^2(r) \, \phi(r) 
	- \int_{\underline{r}-}^r \mu_R(u)  \,\phi(u)\, du\right)  \right|_{\underline{r}-}^{\overline{r}^{SB}-}
    \\
    & - \int_{\underline{r}+}^{\overline{r}^{SB}-} h^{\prime}(r) \, \frac{\partial}{\partial r} \left( \frac{1}{2}\, \sigma_R^2(r) \, \phi(r) 
	- \int_{\underline{r}-}^r \mu_R(u)  \,\phi(u)\, du\right) dr 
    \\    
	& + h^\prime(\overline{r}^{SB}-) \, \left(\int_{\underline{r}+}^{\overline{r}^{SB}-} \mu_R(r) \, \phi(r)\, dr 
    - \kappa\,\overline{r}^{SB}
    %%% + \kappa\,(\underline{r}-\overline{r}^{SB})  
    \, \pi^{SB}(\{ \overline{r}^{SB}\}) \right)
    \\
    = &  - \int_{\underline{r}+}^{\overline{r}^{SB}-} h^{\prime}(r) \, \frac{\partial}{\partial r} \left( \frac{1}{2}\, \sigma_R^2(r) \, \phi(r) 
	- \int_{\underline{r}-}^r \mu_R(u)  \,\phi(u)\, du\right) dr 
    \\    
	& + h^\prime(\overline{r}^{SB}-) \, \left( \frac{1}{2}\, \sigma_R^2(\overline{r}^{SB}-) \, \phi(\overline{r}^{SB}-)  
    - \kappa\, \overline{r}^{SB}
    %%% + \kappa\,(\underline{r}-\overline{r}^{SB})  
    \, \pi^{SB}(\{ \overline{r}^{SB}\}) \right)\,.
\end{align*}
This equation holds for all $h \in \mathcal D$ and therefore
\begin{align*}
    0  & = \frac{\partial}{\partial r} \left( \frac{1}{2}\, \sigma_R^2(r) \, \phi(r) 
	- \int_{\underline{r}-}^r \mu_R(u)  \,\phi(u)\, du\right) \, \quad \text{for all } r \in (\underline{r},\overline{r}^{SB})\, , 
    \\
    0 & = \frac{1}{2}\, \sigma_R^2(\overline{r}^{SB}-) \, \phi(\overline{r}^{SB}-)  
    - \kappa\,\overline{r}^{SB}
    %%% + \kappa\,(\underline{r}-\overline{r}^{SB})  
    \, \pi^{SB}(\{ \overline{r}^{SB}\})
\end{align*}
The solution $\phi$ is of the form 
\begin{align*}
	\phi(r) & = \lambda \, \frac{2\, e^{\int_{\underline{r}}^r \frac{2\,\mu_R(u)}{\sigma_R^2(u)}\,{\rm d}u }}{\sigma_R^2(r)}\, , \text{ for all } r \in (\underline{r},\overline{r}^{SB})\,,
\end{align*}
for some parameter $\lambda \ge 0$, and
\begin{align*}
    0 & = \lambda \,e^{\int_{\underline{r}}^{\overline{r}^{SB}} \frac{2\,\mu_R(u)}{\sigma_R^2(u)}\,{\rm d}u }  
    - \kappa\,\overline{r}^{SB}
    %%% + \kappa\,(\underline{r}-\overline{r}^{SB})  
    \, \pi^{SB}(\{ \overline{r}^{SB}\})\,.
\end{align*}
To pin down $\lambda$, observe that
\begin{align*}
    1  & = \int_{\underline{r}+}^{\overline{r}^{SB}-} \phi(r)\, dr + \pi^{SB}(\{ \overline{r}^{SB}\})
    = \lambda \int_{\underline{r}+}^{\overline{r}^{SB}-}  \frac{2\, e^{\int_{\underline{r}}^r \frac{2\,\mu_R(u)}{\sigma_R^2(u)}\,{\rm d}u }}{\sigma_R^2(r)}\, dr + \pi^{SB}(\{ \overline{r}^{SB}\})\,.
\end{align*}
Eliminate $\pi^{SB}(\{ \overline{r}^{SB}\})$ to determine $\lambda$
\begin{align*}
    0 & = \lambda \,e^{\int_{\underline{r}}^{\overline{r}^{SB}} \frac{2\,\mu_R(u)}{\sigma_R^2(u)}\,{\rm d}u }  
    - \kappa\,\overline{r}^{SB}   
    %%% + \kappa\,(\underline{r}-\overline{r}^{SB}) 
    \left( 1 -\lambda \int_{\underline{r}+}^{\overline{r}^{SB}-}  \frac{2\, e^{\int_{\underline{r}}^r \frac{2\,\mu_R(u)}{\sigma_R^2(u)}\,{\rm d}u }}{\sigma_R^2(r)}\, dr\right)
    \\
    & = \lambda\left( e^{\int_{\underline{r}}^{\overline{r}^{SB}} \frac{2\,\mu_R(u)}{\sigma_R^2(u)}\,{\rm d}u }  
    + \kappa\,\overline{r}^{SB}
    %%% - \kappa\,(\underline{r}-\overline{r}^{SB}) 
    \int_{\underline{r}+}^{\overline{r}^{SB}-}  \frac{2\, e^{\int_{\underline{r}}^r \frac{2\,\mu_R(u)}{\sigma_R^2(u)}\,{\rm d}u }}{\sigma_R^2(r)}\, dr \right) 
    - \kappa\,\overline{r}^{SB}
    %%% + \kappa\,(\underline{r}-\overline{r}^{SB}) 
    \,.
\end{align*}
Solving for $\lambda$ gives
\begin{align*}
    \lambda & = \frac{
    \kappa\,\overline{r}^{SB}
    %%% \kappa\,(\overline{r}^{SB}- \underline{r}) 
    \, e^{-\int_{\underline{r}}^{\overline{r}^{SB}} \frac{2\,\mu_R(u)}{\sigma_R^2(u)}\,{\rm d}u } }{ 1  +
    \kappa\,\overline{r}^{SB} 
    %%% \kappa\,(\overline{r}^{SB} - \underline{r})
    \int_{\underline{r}+}^{\overline{r}^{SB}-}  \frac{2\, e^{-\int_r^{\overline{r}^{SB}} \frac{2\,\mu_R(u)}{\sigma_R^2(u)}\,{\rm d}u }}{\sigma_R^2(r)}\, dr}\,.
\end{align*}
The stationary distribution is then characterized by
\begin{align*}
    \phi(r) & = \frac{
    \kappa\, \overline{r}^{SB}
    %%% \kappa\,(\overline{r}^{SB}- \underline{r}) 
    \,  \frac{2\, e^{-\int_{r}^{\overline{r}^{SB}} \frac{2\,\mu_R(u)}{\sigma_R^2(u)}\,{\rm d}u }}{\sigma_R^2(r)} }{ 1  + 
    \kappa\, \overline{r}^{SB}  
    %%% \kappa\,(\overline{r}^{SB} - \underline{r}) 
    \int_{\underline{r}+}^{\overline{r}^{SB}-}  \frac{2\, e^{-\int_y^{\overline{r}^{SB}} \frac{2\,\mu_R(u)}{\sigma_R^2(u)}\,{\rm d}u }}{\sigma_R^2(y)}\, dy} , \text{ for all } r \in (\underline{r},\overline{r}^{SB})\, ,
    \\
    \pi^{SB}(\{ \overline{r}^{SB}\}) & = \frac{ 1 }{ 1  +
    \kappa\,\overline{r}^{SB} 
    %%% \kappa\,(\overline{r}^{SB} - \underline{r}) 
    \int_{\underline{r}+}^{\overline{r}^{SB}-}  \frac{2\, e^{-\int_r^{\overline{r}^{SB}} \frac{2\,\mu_R(u)}{\sigma_R^2(u)}\,{\rm d}u }}{\sigma_R^2(r)}\, dr} \, . 
\end{align*}
Plugging in the coefficients of $R$, that is, $\mu_R$ and $\sigma_R^2$, gives
\begin{align*}
    \phi(r) & = \frac{ \displaystyle 
    \kappa\,\overline{r}^{SB}  
    %%% \kappa\,(\overline{r}^{SB}- \underline{r})  
    \, \frac{2\, e^{- \frac{\gamma}{\kappa}(\overline{r} - r)  } }{\kappa^2\,\sigma^2\, s^2(r)} }{ \displaystyle 1 + 
    \kappa\,\overline{r}^{SB} 
    %%% \kappa\,(\overline{r}^{SB}- \underline{r}) 
    \, \int_{\underline{r}+}^{\overline{r}-} \frac{2\, e^{- \frac{\gamma}{\kappa}(\overline{r} -u)  }}{\kappa^2\,\sigma^2\, s^2(u)} \,{\rm d}u }\, , \text{ for all } r \in (\underline{r},\overline{r}^{SB})\, ,
    \\
	\pi^{SB}(\{ \overline{r}^{SB}\}) & = \frac{ 1 }{ \displaystyle 1 + 
    \kappa\,\overline{r}^{SB} 
    %%% \kappa\,(\overline{r}^{SB}- \underline{r}) 
    \, \int_{\underline{r}+}^{\overline{r}-} \frac{2\, e^{- \frac{\gamma}{\kappa}(\overline{r} - r)  } }{\kappa^2\,\sigma^2\, s^2(r)} \,{\rm d}r } \,.
\end{align*}
\end{proof}

\end{document}